\documentclass[aps,pre,amsmath,superscriptaddress,reprint,notitlepage,amssymb]{revtex4-1}     

\usepackage{graphicx}
\usepackage{color}
\usepackage{multirow}
\usepackage{setspace}
\usepackage{float}
\usepackage{subfigure}

\pdfoutput=1

\newcommand{\blue}[1]{\color{black} #1}

\newcommand{\overbar}[1]{\mkern 1.5mu\overline{\mkern-1.5mu#1\mkern-1.mu}\mkern 3mu}

\begin{document}
\title{Surface tension of supercooled water nanodroplets from computer simulations}

\author{Shahrazad M.A. Malek}
\affiliation{Department of Physics and Physical Oceanography, Memorial University of Newfoundland, \\St.~John's, Newfoundland A1B 3X7, Canada}

\author{Peter H. Poole}
\affiliation{Department of Physics, St.~Francis Xavier University, Antigonish, NS, B2G 2W5, Canada}

\author{Ivan Saika-Voivod}
\email{saika@mun.ca}
\affiliation{Department of Physics and Physical Oceanography, Memorial University of Newfoundland, \\St.~John's, Newfoundland A1B 3X7, Canada}

\begin{abstract}
We estimate the liquid-vapour surface tension from simulations of TIP4P/2005 water nanodroplets of size $N$=100 to 2880 molecules over a temperature $T$ range of 180~K to 300~K.
We compute the planar surface tension $\gamma_p$, the curvature-dependent surface tension $\gamma_s$, and the Tolman length $\delta$, via
two approaches, one based on the pressure tensor (the ``mechanical route'') and the other on the Laplace pressure (the ``thermodynamic route'').
We find that these two routes give different results for $\gamma_p$, $\gamma_s$ and $\delta$, although in all cases we find that $\delta\ge 0$ and is independent of $T$.  Nonetheless, the $T$ dependence of $\gamma_p$ is consistent between the two routes and with that of Vega and de Miguel [{\it J.~Chem.~Phys.} {\bf 126}, 154707 (2007)] down to the crossing of the Widom line at 230~K for ambient pressure.  Below 230~K, $\gamma_p$ rises more rapidly on cooling  than predicted from behavior for $T\ge 300$~K.  We show that the increase in $\gamma_p$ at low $T$ is correlated to the emergence of a well-structured random tetrahedral network in our nanodroplet cores, and thus that the surface tension can be used as a probe to detect behavior associated with the proposed liquid-liquid phase transition in supercooled water.
\end{abstract}
\date{\today}
\maketitle

\section{Introduction}
\label{Intro}

Microscopic and nanoscopic water droplets are of interest in many important research areas, such as the Earth's climate~\cite{baker,wil}, biological applications~\cite{Ohno}, interstellar space~\cite{Tachibana}, and numerous other systems~\cite{klemp}.  In all of these areas, the surface tension of the liquid-vapour interface of the water droplet is a central physical property for understanding and predicting droplet behavior.  For example, the surface tension is crucial for estimating the nucleation rate of liquid from the vapour using classical nucleation theory~\cite{kulmala,Debenedetti-book}.

The surface tension is also the origin of the pressure difference that arises between the interior and exterior of a liquid droplet, as quantified by the Young-Laplace equation~\cite{young1805,laplace1805},
\begin{equation}
\Delta P = \frac{2\gamma_s}{R_s}.
\label{young-laplace}
\end{equation}
Here, $\Delta P=P_l-P_v$, where $P_l$ and $P_v$ are the respective pressures of the liquid interior and vapour exterior, and $\gamma_s$ is the surface tension of the curved interface.  {\blue $R_s$ is the radius of the so-called ``surface of tension"~\cite{Rowlinson1982}.}
For macroscopic droplets, the width of the molecular interface is negligible compared to the droplet dimensions, and $R_s$ is simply the radius of the droplet.  
However, for nanoscale droplets, the interfacial width is significant compared to the size of the droplet itself, and various definitions for the radius of the droplet are possible.

It has long been understood that the surface tension of a curved interface deviates from that of a planar interface. 
For a curved surface, such as that of a droplet, the Tolman length $\delta$ quantifies how
$\gamma_s$ deviates from the planar surface tension $\gamma_p$ as a function of $R_s$, via the expression~\cite{Tolman1949},
\begin{equation}
\gamma_s = \frac{\gamma_p}{\left(1+2\delta/R_s\right)}.
\label{tolman}
\end{equation}
The magnitude of $\delta$ is generally found to be 10-20\% of the molecular diameter.  

However, the sign of $\delta$ is a subject of continuing debate~\cite{Blokhuis2009}. 
While modeling on the basis of classical density
functional theory has predicted negative values of $\delta$ for liquid Lennard-Jones (LJ) droplets~\cite{Blokhuis2013,reguera2015}, simulations of droplets have estimated both negative and positive values of $\delta$. For example, Yan, {\it et al.}~\cite{Yan2016} performed MD simulations of liquid argon nanodroplets with sizes ranging from 800 to 2000 atoms at 78~K, as modelled using the LJ potential. They evaluated the pressure tensor, and using the Young-Laplace equation they concluded that $\delta$ is positive for LJ nanodroplets. However, Giessen and Blokhuis~\cite{Blokhuis2009} estimated a negative value of $\delta$ for LJ nanodroplets. 

A similar disagreement regarding the magnitude and sign of $\delta$ appears in water simulations. Leong and Wang~\cite{Leong2018} performed MD simulations using the BLYPSP-4F water potential~\cite{BLYP} on nanoscale  droplets with radii varying between 2 and 8~nm at temperature $T=298$~K. Using an
empirical correlation between the pressure and density, they estimated $\delta=-0.048$~nm.  
A similar value for $\delta$ was obtained by measuring the free energy of droplet mitosis in a study by Joswiak, {\it et al.}~\cite{Joswiak} for the mW model of water~\cite{mW}.
On the other hand, Lau, {\it et al.}~\cite{Lau2015} used a test-area method and obtained a positive value of $\delta$ for the TIP4P/2005 model of water~\cite{vega2005}.  
Simulation studies of cavitation for TIP4P/2005 find relatively large positive values of 
$\delta$ for vapour bubbles with magnitudes in the range of 0.12 to 0.195~nm~\cite{menzl2016, min2019}.  This result implies that for a TIP4P/2005 water droplet of the same size, $\delta$ should be of similar magnitude, but negative.  It is evident from this recent work that disagreement exists on both the magnitude and sign of $\delta$, even when the same water model is used.

The variation of the surface tension with  $T$ for deeply supercooled water has also been investigated, in particular as a way to test for evidence of a possible liquid-liquid phase transition (LLPT) in supercooled water~\cite{Poole1992-Phase}.   Theoretical studies have shown that if a LLPT occurs, then at low $T$ the surface tension should increase faster with decreasing $T$ than is expected otherwise~\cite{feeney,hruby2004,hruby2005}.  Some computer simulations studies are consistent with this behavior~\cite{lu2006a,lu2006b} while others are not~\cite{chen,vier}. 
Recent careful experiments by Hruby and coworkers do not find evidence for a change in the $T$ dependence of the surface tension for $T$ as low as $-26$~$^\circ$C~\cite{hruby2014,hruby2015,hruby2017}.  However, it is possible that the anomalous increase in the surface tension will only be observed for $T$ below the Widom line that is associated with the LLPT, a range of $T$ that 
has only recently begun to be probed in experiments~\cite{kim2017}.

The sign of $\delta$ determines whether $\gamma_s$ decreases or increases with $R_s$. For a positive $\delta$, $\gamma_s$ decreases as $R_s$ decreases. Moreover, $\delta$ relates the equimolar radius $R_e$ and $R_s$~\cite{Tolman1949},
\begin{equation}
\delta=R_e-R_s,
\label{delta}
\end{equation}
where $R_e$ is the radius of a sphere that has a uniform density equal to that of the interior part of the droplet and that has the same number of molecules as the droplet.
Since determining $R_e$ is more straightforward than determining $R_s$, we can rewrite Eqs.~\ref{young-laplace} and~\ref{tolman} in terms of $R_e$,
\begin{equation}
\Delta P =\frac{2\gamma_p}{R_e}\left(\frac{1}{1+\delta/R_e}\right),
\label{young-laplaceRe}
\end{equation}
or in the form,
\begin{equation}
\frac{2}{\Delta P R_e} = \frac{1}{\gamma_p}\left(1+\delta/R_e\right),
\label{young-laplaceRev2}
\end{equation}
and 
\begin{equation}
\gamma_s = \gamma_p\frac{R_e-\delta}{R_e+\delta}.
\label{tolmanRe}
\end{equation}

The above equations provide the basis for a procedure to find $\gamma_p$, $\gamma_s$ and $\delta$, 
which following past practise we refer to here as the ``thermodynamic route"~\cite{Thompson1984}.  As we will see below, computer simulations of water nanodroplets allow us to directly estimate $\Delta P$ and $R_e$.  If we obtain $\Delta P$ and $R_e$ for a range of droplet sizes at fixed $T$, we can use Eq.~\ref{young-laplaceRev2} to estimate $\gamma_p$ and $\delta$ by curve fitting.  From $\delta$ an estimate of $R_s$ is obtained from Eq.~\ref{delta}, and so an estimate of $\gamma_s$ can be computed using Eq.~\ref{young-laplace}.

Aside from the Laplace equation, Rowlinson and Widom proposed a model to derive $\gamma_s$ from the tangential and normal components of the pressure tensor as functions of the radial distance $r$ from the centre of mass of a droplet, $P_T(r)$ and $P_N(r)$~\cite{Rowlinson1982}. The model assumes two homogeneous fluid phases, with homogeneous pressures $P^\alpha$ and $P^\beta$ far from the interface, and an inhomogeneous interface between them.  Under the model assumption that the surface tension acts at a single value of $r=R_s$, the mechanical requirements for static equilibrium, i.e. force and torque balance, yield,
\begin{eqnarray}
\gamma_s &=& \int_0^\infty\left(\frac{r}{R_s}\right)\left[P^{\alpha,\beta}(r;R_s)-P_T(r)\right]dr, \label{gammasforce} \\
&=& \int_0^\infty\left(\frac{r}{R_s}\right)^2\left[P^{\alpha,\beta}(r;R_s)-P_T(r)\right]dr,
\label{gammastourqe}
\end{eqnarray}
where $P^{\alpha,\beta}(r;R_s)$ is $P^\alpha$ for $r<R_s$ and $P^\beta$ for $r>R_s$. These equations in turn give an expression for $R_s$,
\begin{equation}
R_s = \frac{\int_0^\infty r^2\left[P^{\alpha,\beta}(r;R_s)-P_T(r)\right]dr}{\int_0^\infty r\left[P^{\alpha,\beta}(r;R_s)-P_T(r)\right]dr}.
\label{Rsnumerical}
\end{equation}
With the assumption that the two phases are homogeneous, we can assume that $P^\alpha=P_l$ and $P^\beta=P_v$. Since $P^{\alpha,\beta}(r;R_s)$ depends on $R_s$, Eq.~\ref{Rsnumerical} must be evaluated numerically.


From the condition of mechanical stability, $\nabla\cdot {\bf P}=0$, it can be shown that,
\begin{equation}
\int_0^\infty r^2\left[P^{\alpha,\beta}(r;R_s)-P_N(r)\right]dr=0,
\label{PnIdentity}
\end{equation} 
and hence $\gamma_s$ can be obtained using the $P_N(r)$ component of the pressure in Eqs.~\ref{gammasforce} and~\ref{gammastourqe}, yielding,\begin{eqnarray}
\gamma_s 
&=& \int_0^\infty\left(\frac{r}{R_s}\right)^2\left[P_N(r)-P_T(r)\right]dr,
\label{gammastourqePn}
\end{eqnarray}
and~\cite{Rowlinson1982},
\begin{equation}
R_s = \frac{\int_0^\infty r^2\left[P_N(r)-P_T(r)\right]dr}{\int_0^\infty r\left[P_N(r)-P_T(r)\right]dr}.
\label{RsnumericalPn}
\end{equation}


We can also find $\gamma_s$ while avoiding the need for $R_s$ by combining Eqs.~\ref{young-laplace} and~\ref{gammastourqePn},
\begin{eqnarray}
\gamma_s^3 &=& \frac{(P_l-P_v)^2}{4}\int_0^\infty r^2\left[P_N(r)-P_T(r)\right]dr \label{gammas3}
\end{eqnarray}

Eqs.~\ref{gammasforce}-\ref{gammas3} provide an alternative pathway, referred to as the ``mechanical route"~\cite{Thompson1984}, 
to compute $\gamma_s$ and $R_s$, as well as $\gamma_p$ and $\delta$.
First, $P_N(r)$ and $P_T(r)$ are calculated from simulations of nanodroplets, with Eqs.~\ref{gammasforce}--\ref{gammas3} yielding values for $\gamma_s$ and $R_s$. Estimates for $\gamma_p$ and $\delta$ are then obtained through Eqs.~\ref{tolman} and~\ref{delta}.

%

In this study, we use the TIP4P/2005 model to simulate water nanodroplets over a wide range of temperatures and sizes and determine $\gamma_s$, $\gamma_p$ and $\delta$ using both the thermodynamic and mechanical routes.
In Section II, we provide details of our simulations. 
In Section III-V, we show that the thermodynamic and mechanical routes give different results for the surface tension and Tolman length for water.  Despite these differences, we find that all methods demonstrate that $\gamma_p$ 
increases more rapidly upon cooling through the Widom line temperature for TIP4P/2005.  In Section VI, we show that the rapid increase in $\gamma_p$ at low $T$ is consistent with a crossover within the core of our nanodroplets from the high density liquid phase (HDL) to the low density liquid phase (LDL) of the LLPT.
We  present a discussion and our conclusions in Section VII.





\section{Simulations}

We recently studied the thermodynamic and structural properties of simulated water nanodroplets ranging in size from $N=$ 100 to 2880  molecules, over a $T$ range of 180 to 300~K~\cite{NATURECOMM}, with molecules interacting through the TIP4P/2005 model~\cite{vega2005}.  The same data set is used in the present study.  We summarize the simulation details below for the reader's convenience.
 

We carry out the simulations in the canonical ensemble -- constant $N$, volume $V$, and $T$. The droplets are located in a periodic cubic box of side length $L$ that increases with $N$ and ranges from 10 to 20~nm.
We ensure the box is large enough to avoid any direct interaction between the water droplet and its periodic images, and small enough to ensure that at most only a few molecules are in the vapour phase. We use a potential cutoff of $L$/2, ensuring that all molecules in the droplet interact without truncation of the potential.  
We use Gromacs v4.6.1~\cite{GROMACS} to carry out our molecular dynamics (MD) simulations. We hold the temperature constant using the Nos{\'e}-Hoover thermostat with time constant 0.1~ps. The equations of motion are integrated with the leap-frog algorithm with a time step of 2~fs.


The data set is generated from two kinds of MD runs: conventional ``single long runs" (SLR), and using a ``swarm relaxation'' method (SWRM)~\cite{swarm}.  For droplet sizes $N=100$, 200, 360, 776, 1100, 1440, and 2880, we use SLRs. For $N=1440$ and 2880, we start our simulations by placing $N$ molecules randomly within the simulation box, and run long enough for the molecules to condense into a single droplet. We harvest an equilibrated $N=1440$ configuration, and progressively remove molecules from the droplet surface to obtain starting configurations for the other droplet sizes.  The slowest relaxation times are approximately 12~ns, and our longest post-equilibration simulations last 2.8~$\mu$s.

For droplet sizes $N=205$, 301, 405, 512, 614, and 729, we use SWRM. To generate initial configurations for each of these droplet sizes, we first remove molecules from the surface of an equilibrated $N=2880$ configuration to obtain the desired size. We first conduct SLRs for each size at $T=200$~K for not less than 350~ns. We then take the last configuration of each run and randomize the velocities using the Maxwell-Boltzmann distribution at $T=220$~K to generate $M$ different configurations, which are used to initiate our swarm relaxation runs. We determine the relaxation time $\tau_s$  for a swarm ensemble from the potential energy autocorrelation function of the system.  See Ref.~\cite{swarm} for details. The final equilibrated $M$ configurations of the ensemble are then used to  initiate an ensemble of runs at $T=200$~K. Similarly, we take the final equilibrated $M$ configurations of the ensemble at $T=200$~K to start a swarm ensemble at $T=180$~K.   

Additionally, we carry out simulations for bulk liquid TIP4P/2005 with $T$ varying from $300$ to $180$~K.  We simulate 360 molecules with density varying approximately between $0.96$ and $1.12$~g/cm$^3$ using the protocols described in Ref.~\cite{saika2013}.

\begin{figure}
\centerline{\includegraphics[scale=0.35]{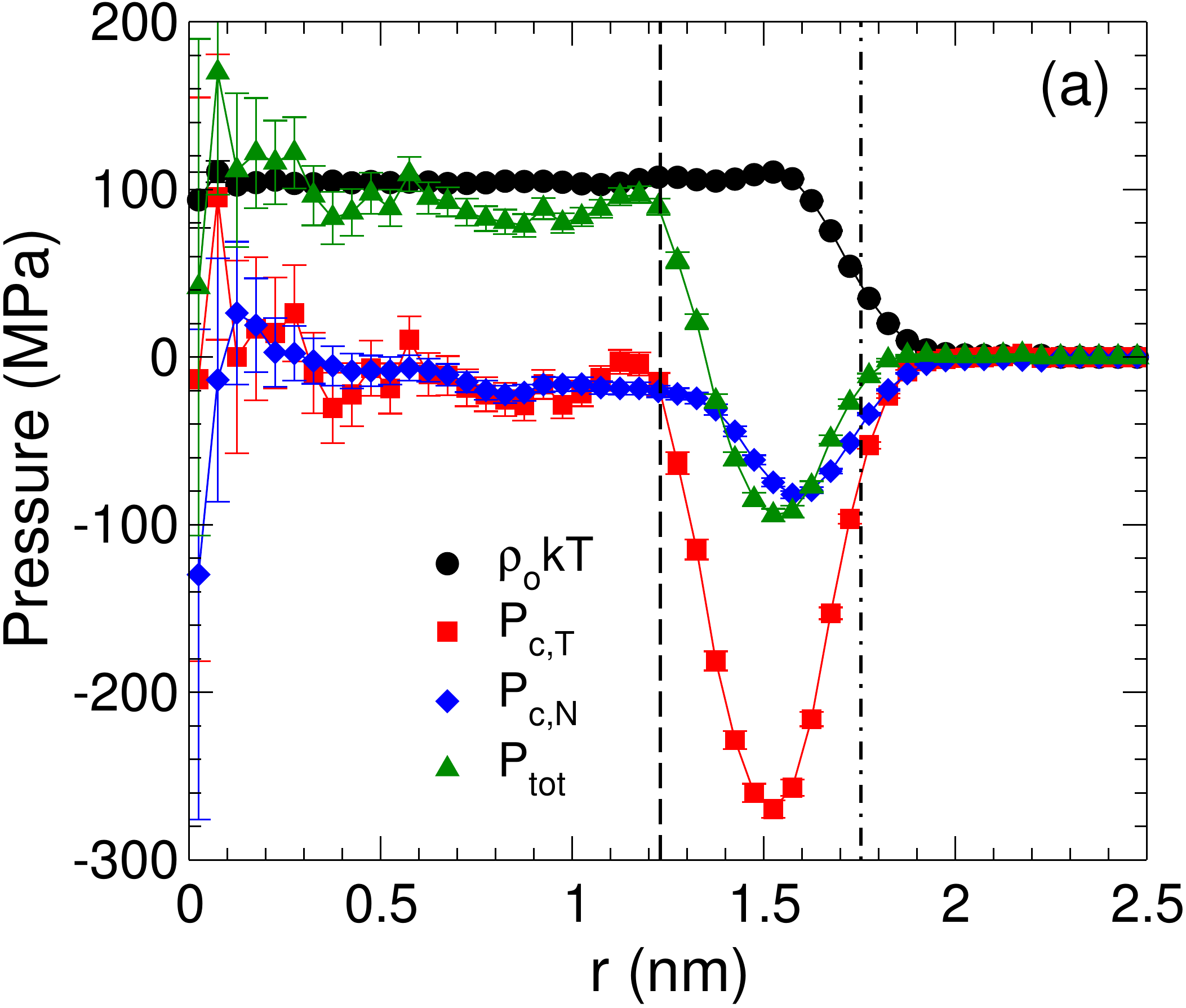}}
\centerline{\includegraphics[scale=0.35]{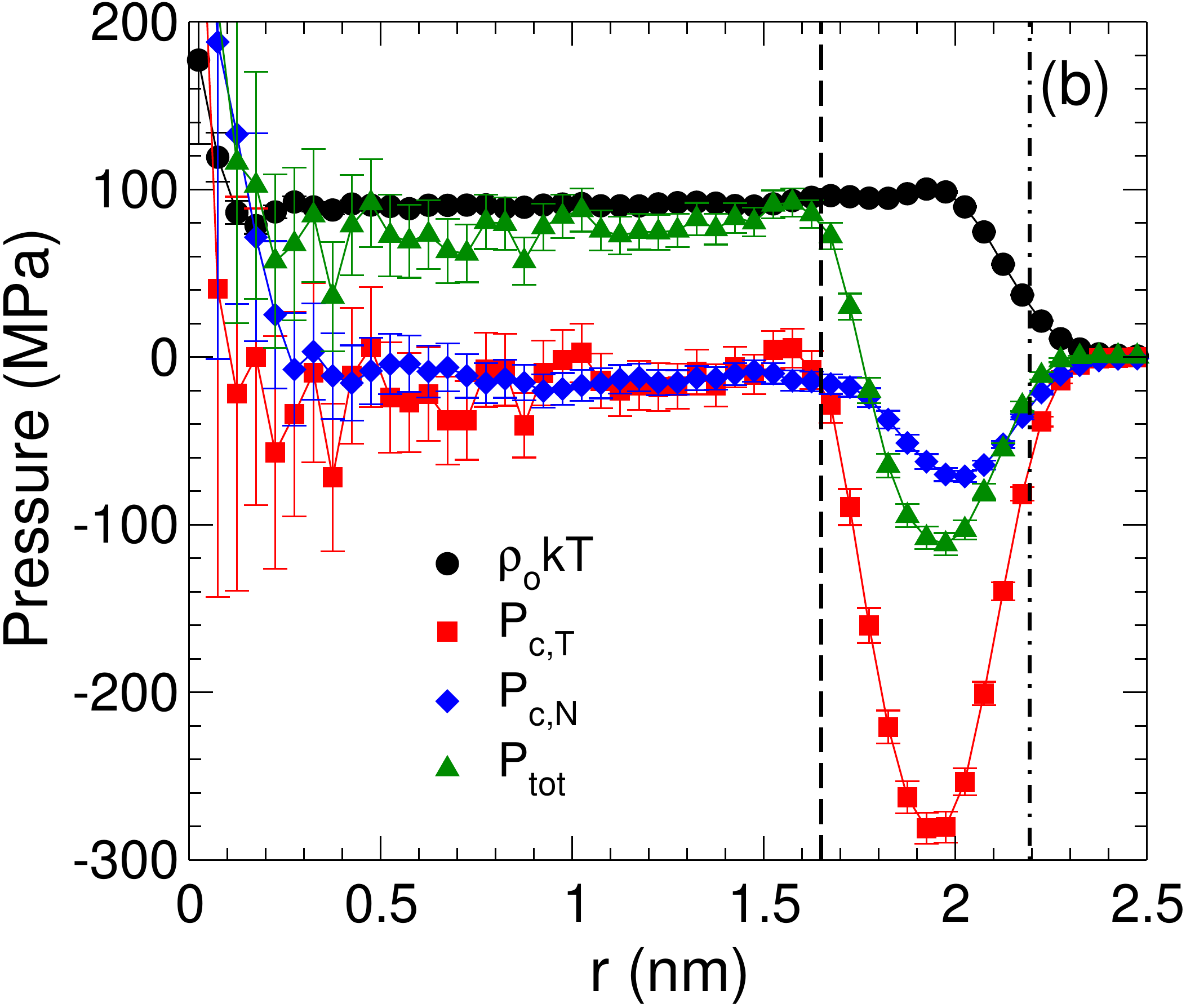}}
\caption[Contributions to the pressure inside water nanodroplets as a function of $r$.]{
Contributions to the pressure inside water nanodroplets as a function of $r$, for (a) $N=776$ and $T=220$~K, and (b) $N=1440$ and $T=200$~K. Vertical lines identify $r=R_L$ (dashed) and $r=R_e$ (dot-dashed). 
}
\label{pressprofiles}
\end{figure}

The mechanical route to
finding the surface tension of a droplet requires the determination of both $P_T(r)$ and $P_N(r)$.  We compute kinetic and configurational contributions to the pressure inside our droplets; see Ref.~\cite{malek2} for details {\blue on applying to TIP4P/2005 a coarse-grained method~\cite{Ikeshoji2011} based on the Irving-Kirkwood~\cite{IrvingKirkwood} choice of contour in defining the microscopic pressure}. Fig.~\ref{pressprofiles} shows all contributions to the pressure for two example cases. We define $R_L$ such that the configurational contributions to the normal and tangential pressures, ${P}_{c,N}$ and ${P}_{c,T}$ respectively, are equal to each other within error for  $r < R_L$ (dashed line in Fig.~\ref{pressprofiles}), noting that they differ near the surface. To define the pressure in the interior of the droplets $P_L$, we average the total (isotropic) pressure $P_{tot}(r) = {P}_{c,N}(r)/3 + 2 {P}_{c,T}(r) /3 + \rho_\circ(r) k_{\rm B} T$ over the spherical volume of radius $R_L$, where $\rho_\circ(r)$ is the local number density.

All error bars reported in this work indicate one standard deviation in the mean.

\begin{figure}
\centerline{\includegraphics[scale=0.35]{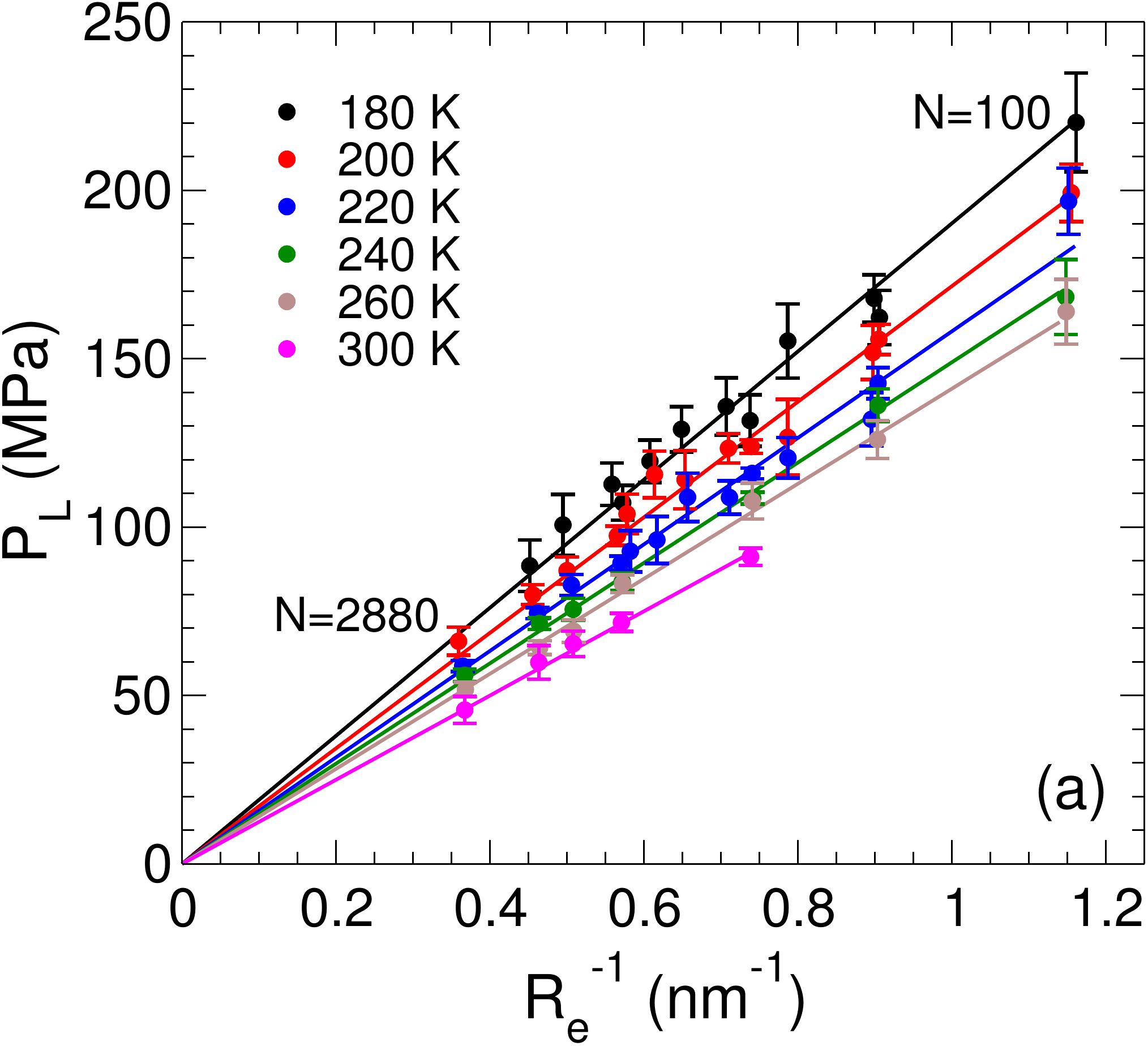}}
\centerline{\includegraphics[scale=0.35]{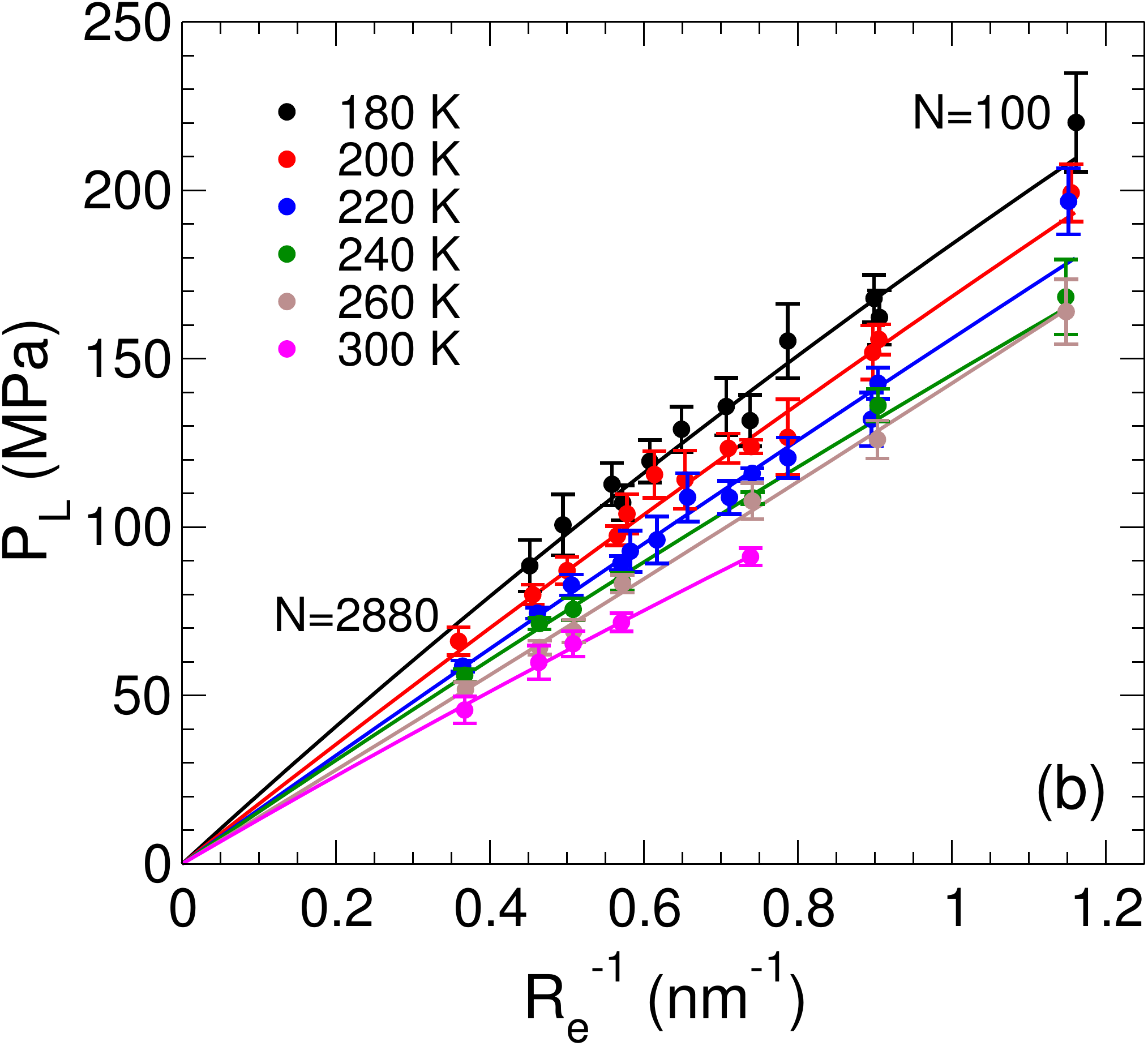}}
\caption[Isotherms of $P_L$ as a function of $R_e^{-1}$.]{
Isotherms of $P_L$ as a function of $R_e^{-1}$. Along each isotherm, $N$ decreases with $R_e$. (a) The straight lines are one-parameter fits to Eq.~\ref{young-laplaceRe} with the assumption that $\delta=0$. (b) The curves are two-parameter fits to Eq.~\ref{young-laplaceRe}. 
}
\label{PLaplacefig}
\end{figure}

\begin{figure}
  \centering
\includegraphics[scale=0.35]{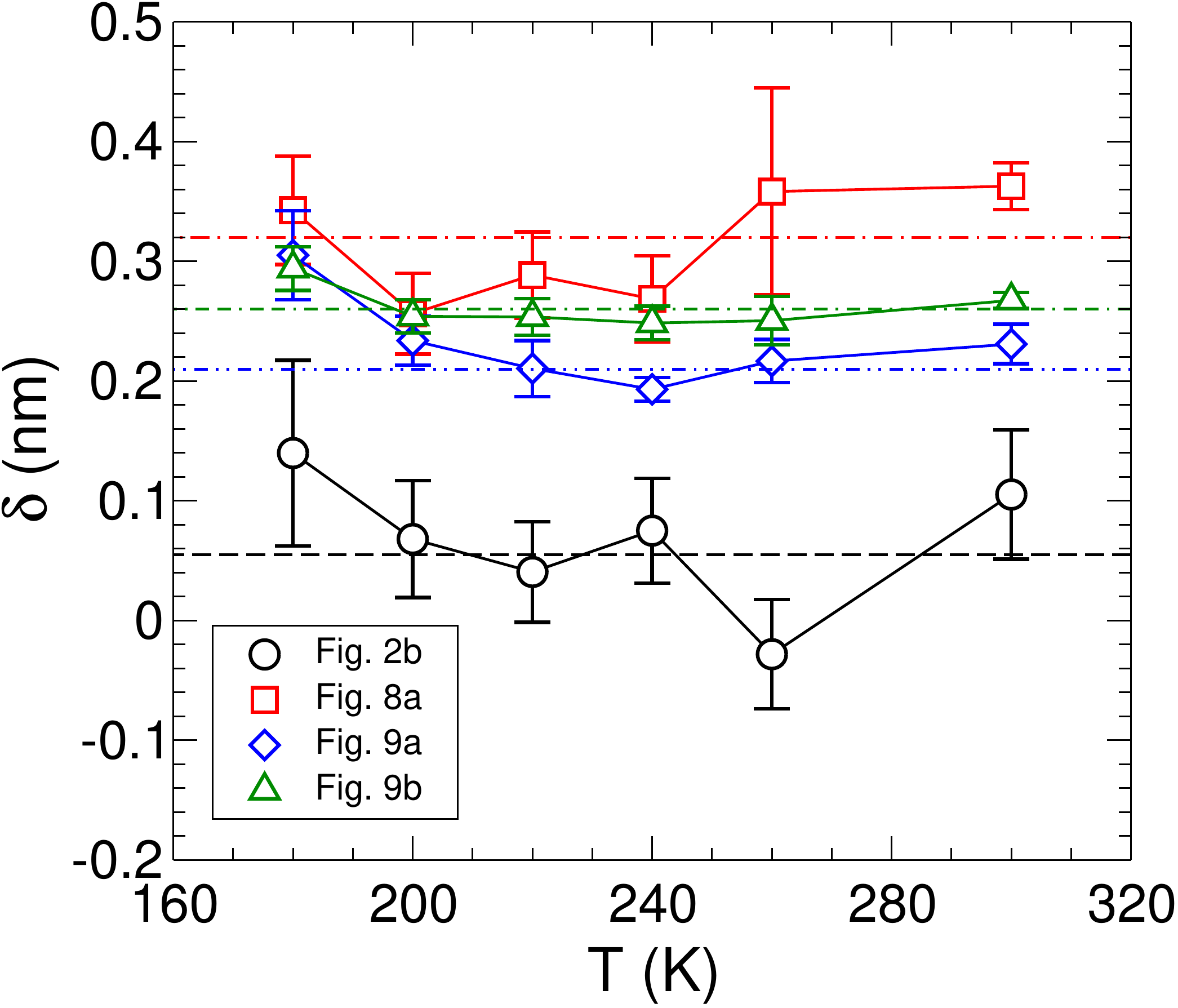}
\caption[Tolman length $\delta$ as a function of $T$ obtained by different means.]{
Tolman length $\delta$ as a function of $T$ obtained by different means:
fits of $P_L(R_e)$ to Eq.~\ref{young-laplaceRe} shown in Fig.~\ref{PLaplacefig}b (black circles) with average value of $0.055\pm 0.021$~nm (dashed line);
fits of $\gamma_s(R_s)$ to Eq.~\ref{tolman} shown in Fig.~\ref{gammaSfig}a (red squares) with average value of $0.32\pm 0.02$~nm (dot-dash);
fits of $\gamma_s(R_s)$ to Eq.~\ref{tolman} shown in Fig.~\ref{gammaS3fig}a (blue diamonds) with average value of $0.21\pm 0.01$~nm (dot-dot-dash); and
fits of $\gamma_s(R_e)$ to Eq.~\ref{tolmanRe} shown in Fig.~\ref{gammaS3fig}b (green triangles) with average value of $0.26\pm 0.005$~nm (dot-dash-dash).
%
%
The top three curves are from the mechanical route, while the bottom curve (black circles) is from the thermodynamic route. 
}
\label{deltaPL}
\end{figure}

\begin{figure}
\centerline{\includegraphics[scale=0.35]{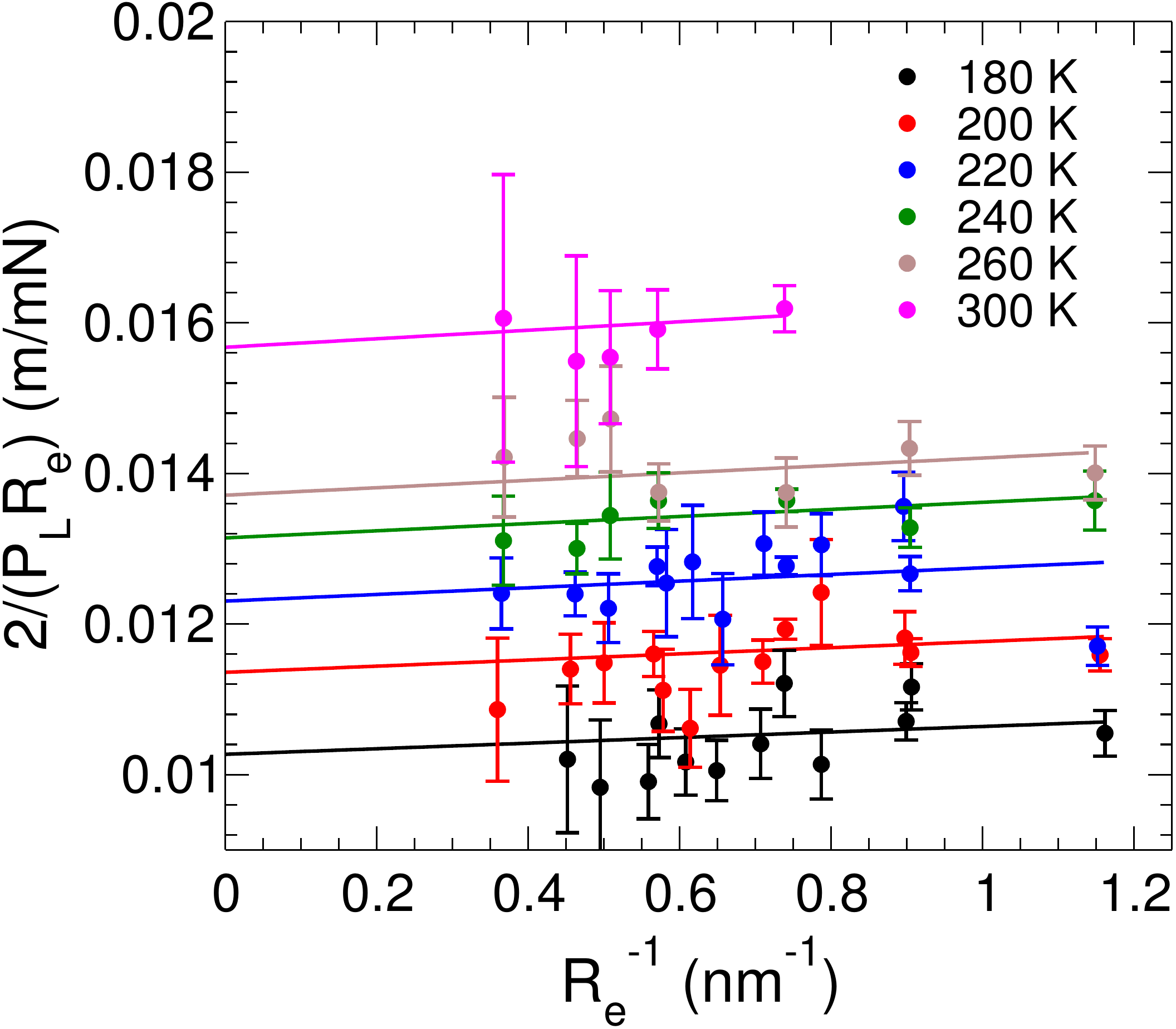}}
\caption{
Isotherms of $2/(P_LR_e)$ as a function of $R_e^{-1}$. Along each isotherm, $N$ decreases with $R_e$. The straight lines are fits to Eq.~\ref{young-laplaceRev2}, where $\delta=0.056$~nm is a global fit parameter. 
}
\label{PREfig}
\end{figure}

\begin{figure}
  \centering
  \includegraphics[scale=0.35]{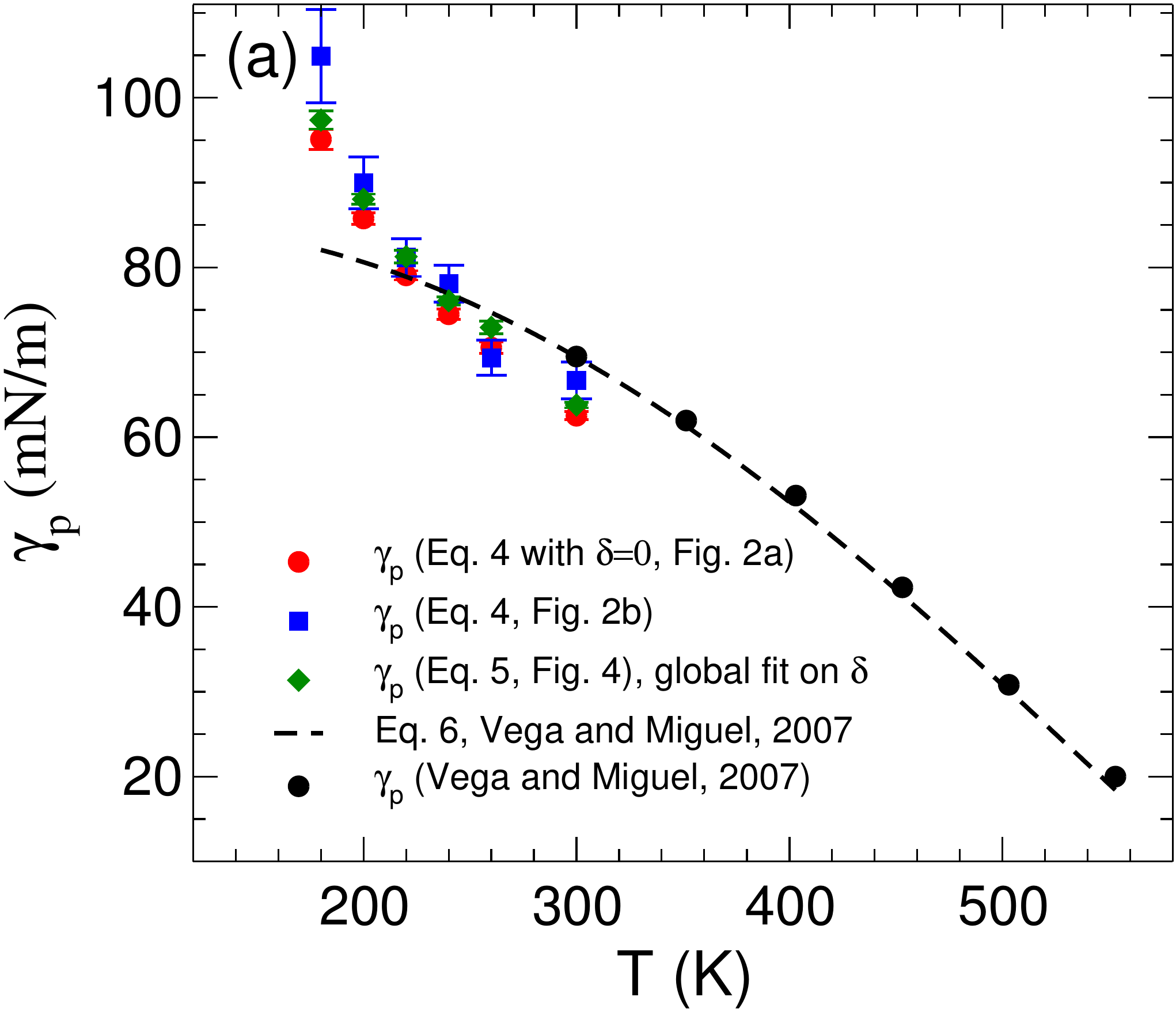}
  \includegraphics[scale=0.35]{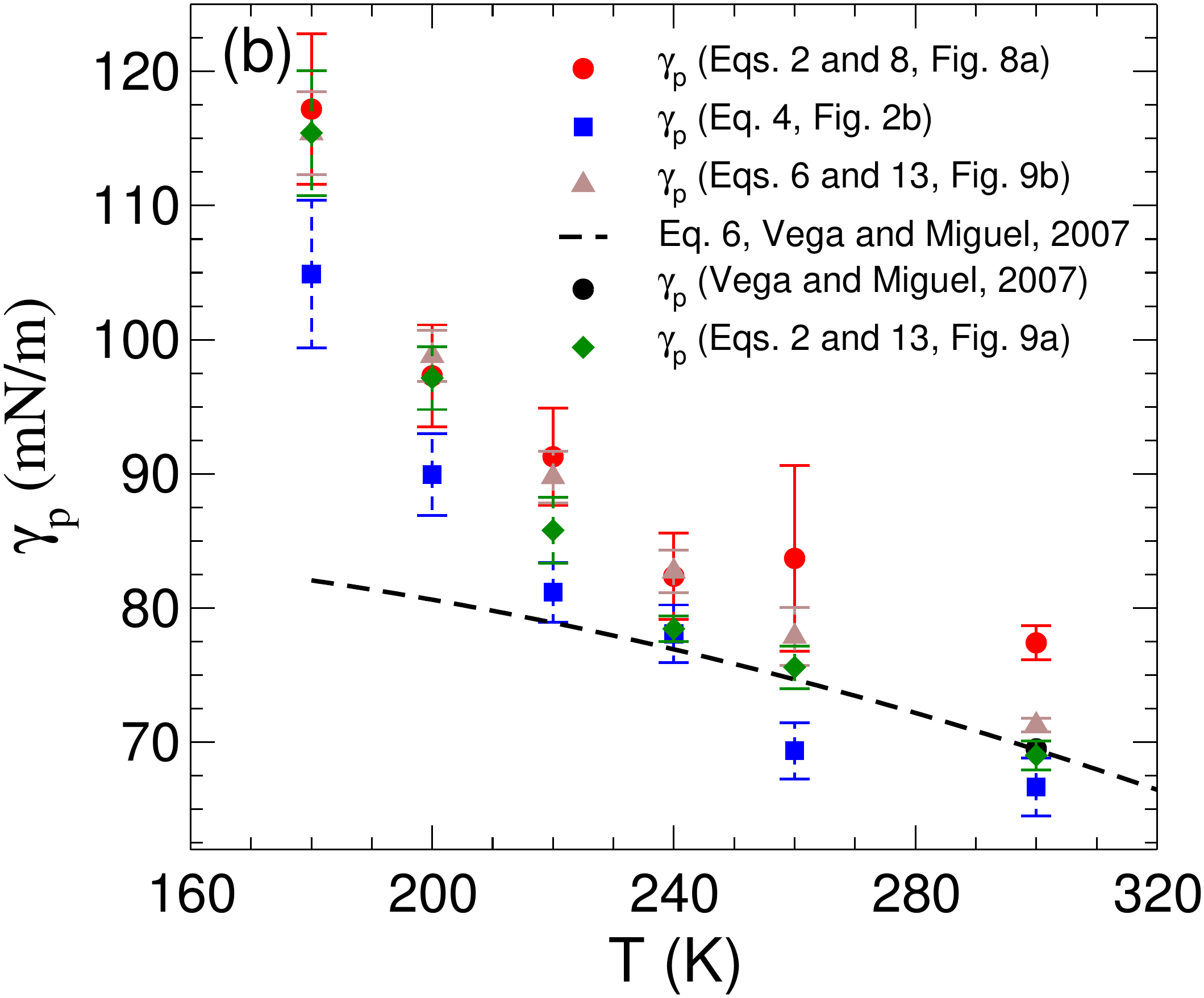}
\caption{The variation of planar surface tension $\gamma_p$ with $T$. (a) $\gamma_p$ via the thermodynamic route obtained from the fits in Fig.~\ref{PLaplacefig}a (red circles), Fig.~\ref{PLaplacefig}b (blue squares), Fig.~\ref{PREfig} (green diamonds). (b) $\gamma_p$ via the mechanical route obtained from the fits in Fig.~\ref{gammaSfig}a (red circles), Fig.~\ref{gammaS3fig}a (green diamonds), and Fig.~\ref{gammaS3fig}b (brown triangles).  Thermodynamic route results from Fig.~\ref{PLaplacefig}b (blue squares) are added for comparison.}
\label{gammapfig}
\end{figure}

\begin{figure}
\centering
\includegraphics[scale=0.35]{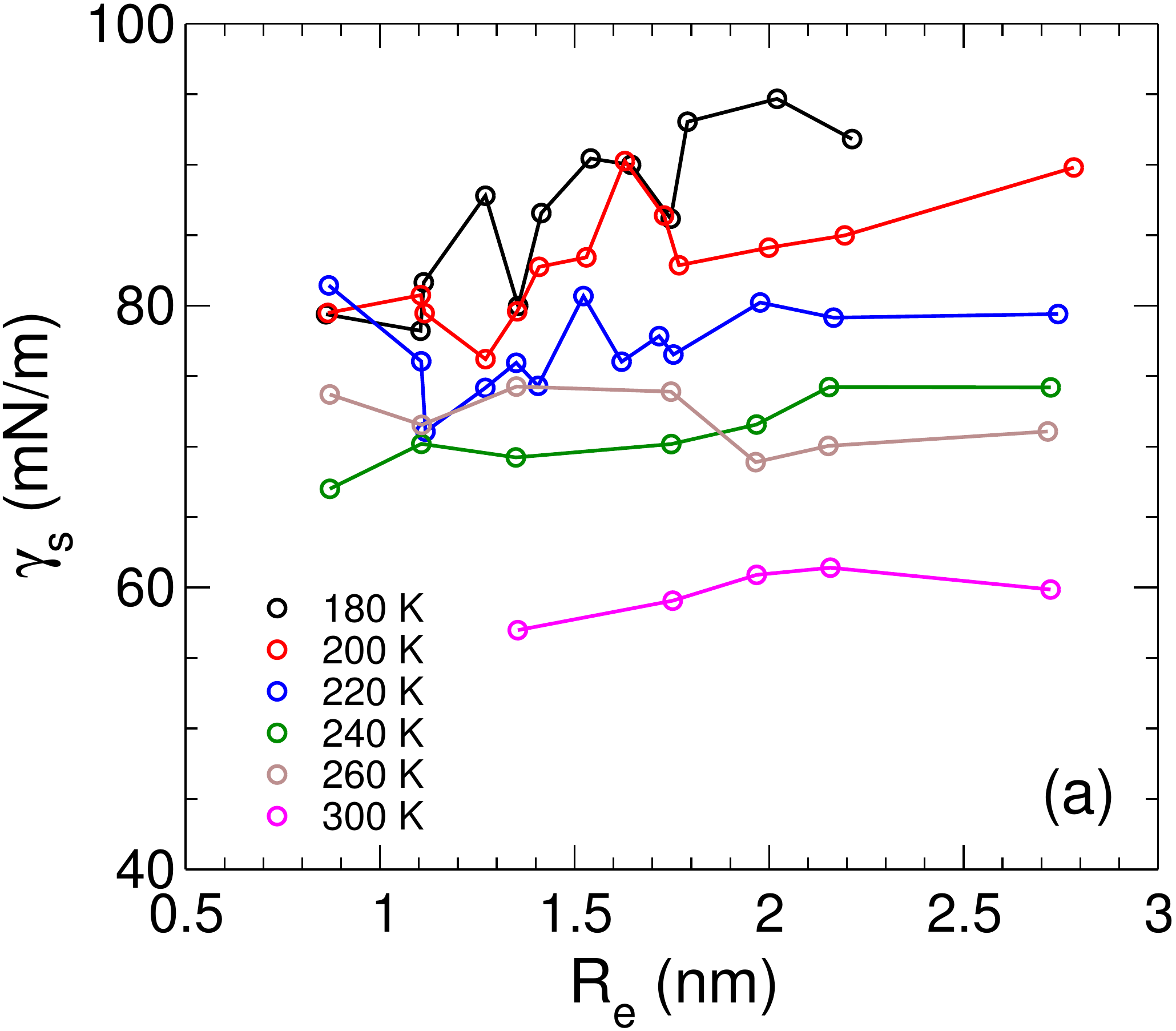}
\includegraphics[scale=0.35]{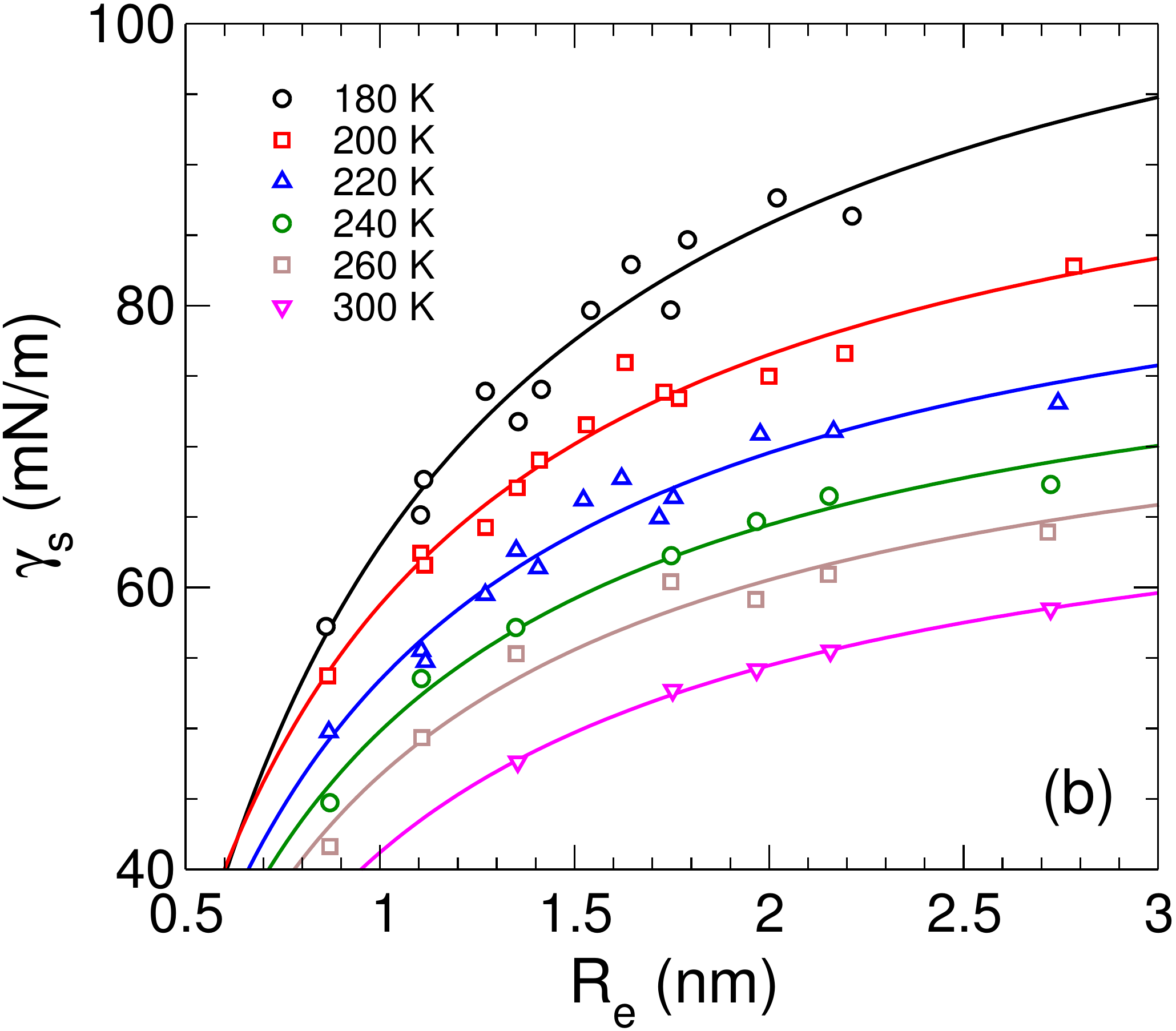}
\caption{
Isotherms of $\gamma_s$ as a function of $R_e$. 
(a) $\gamma_s$ obtained from the thermodynamic route, $\gamma_s=P_L \, (R_e-\delta)/2$. 
(b) $\gamma_s$ obtained from the mechanical route through Eq.~\ref{gammas3}. Curves are fits to Eq.~\ref{tolmanRe}.
}
\label{gammap-allTfig}
\end{figure}

\begin{figure}
\centerline{\includegraphics[scale=0.35]{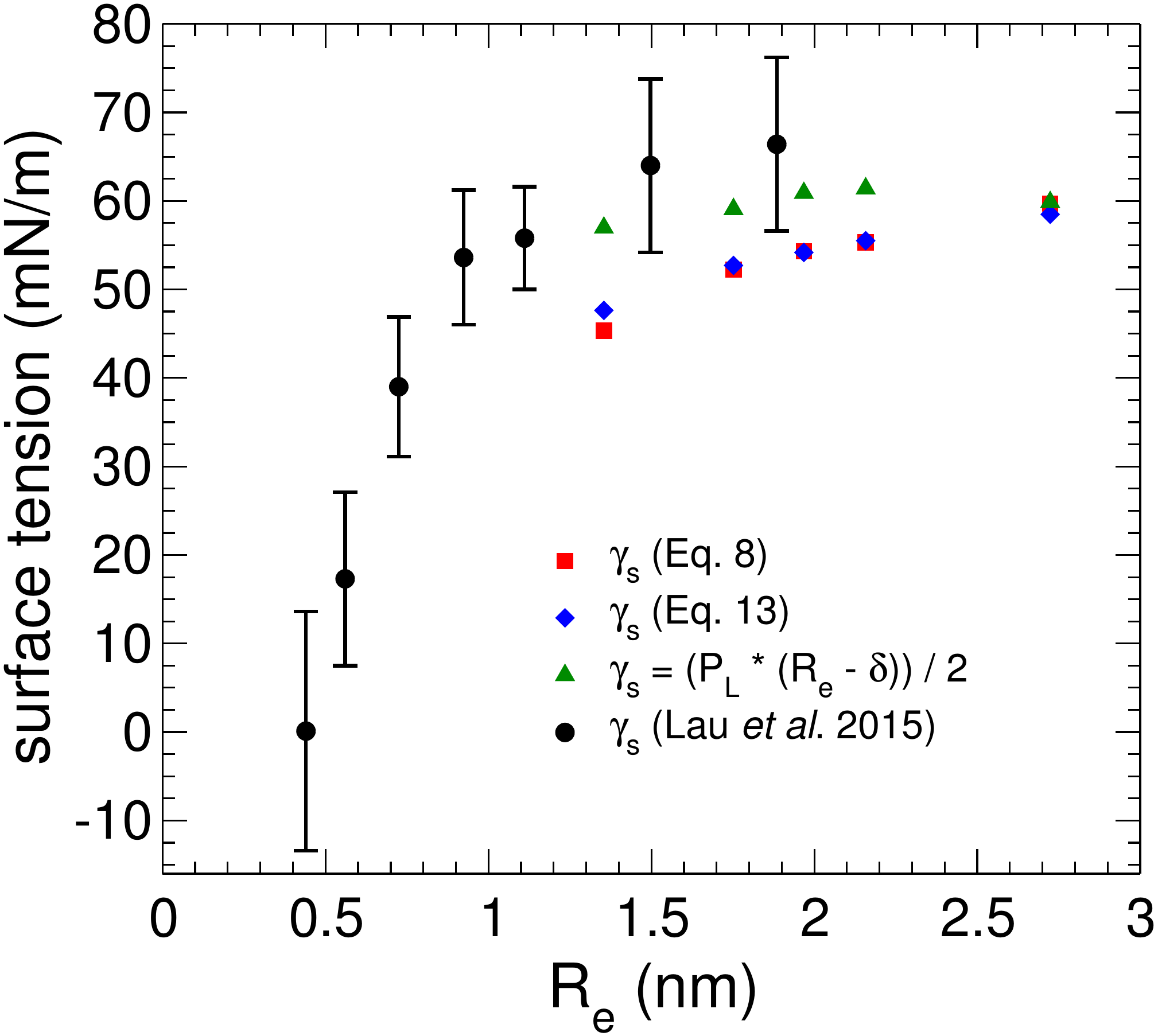}}
\caption{
$\gamma_s$ as a function of $R_e$ at $T=300$~K from Eq.~\ref{gammastourqe} (red squares); Eq.~\ref{gammas3} (blue diamonds);
using $\gamma_s=P_L \, (R_e-\delta)/2$ with $\delta=0.105$~nm obtained from Fig.~\ref{PLaplacefig}b
(green triangles); and from {Lau, \it et al.}~\cite{Lau2015} at $T=293$~K (black circles).
}
\label{gammap-300Kfig}
\end{figure}

\begin{figure}[!htb]
\centerline{\includegraphics[scale=0.35]{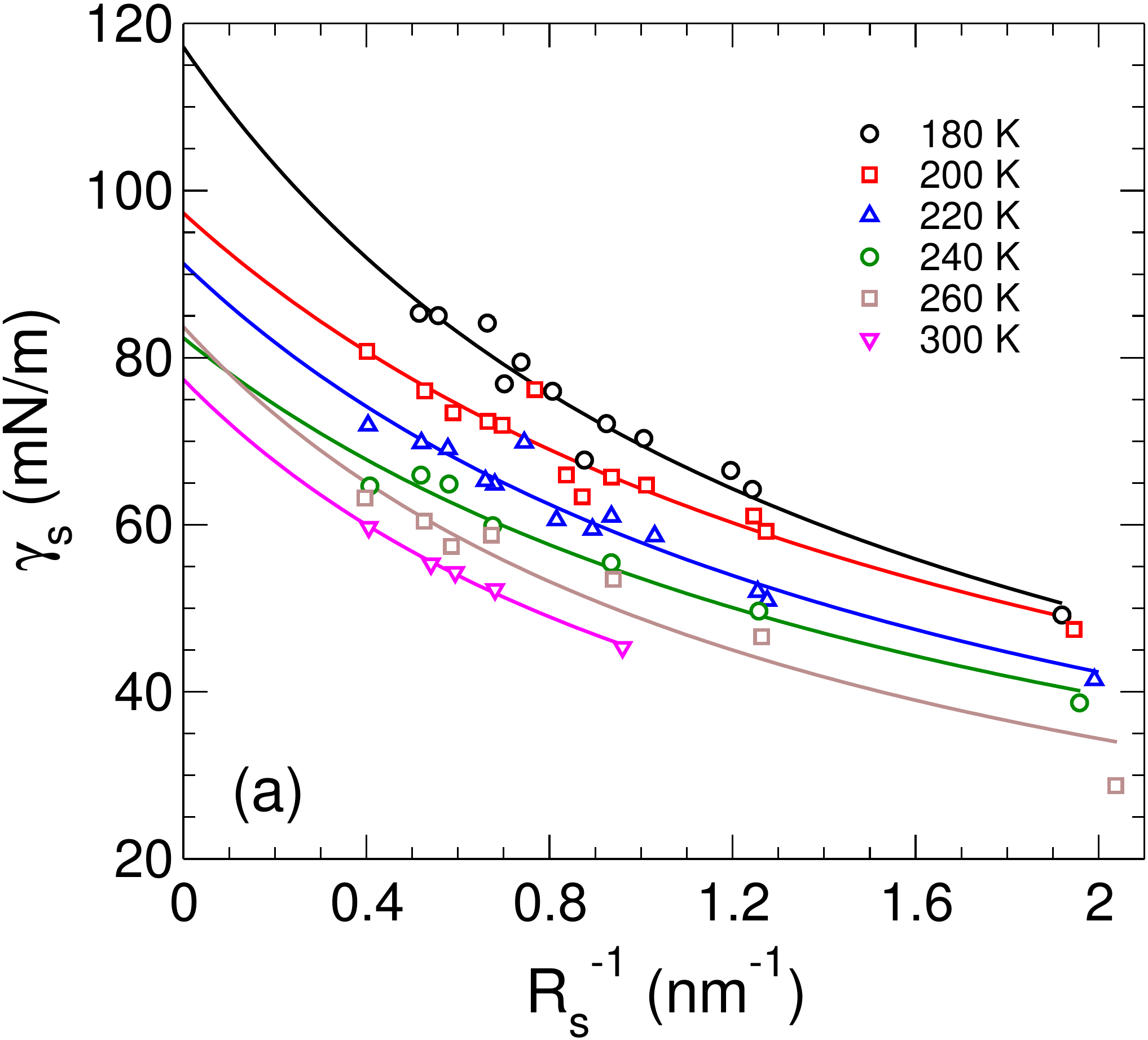}}
\centerline{\includegraphics[scale=0.35]{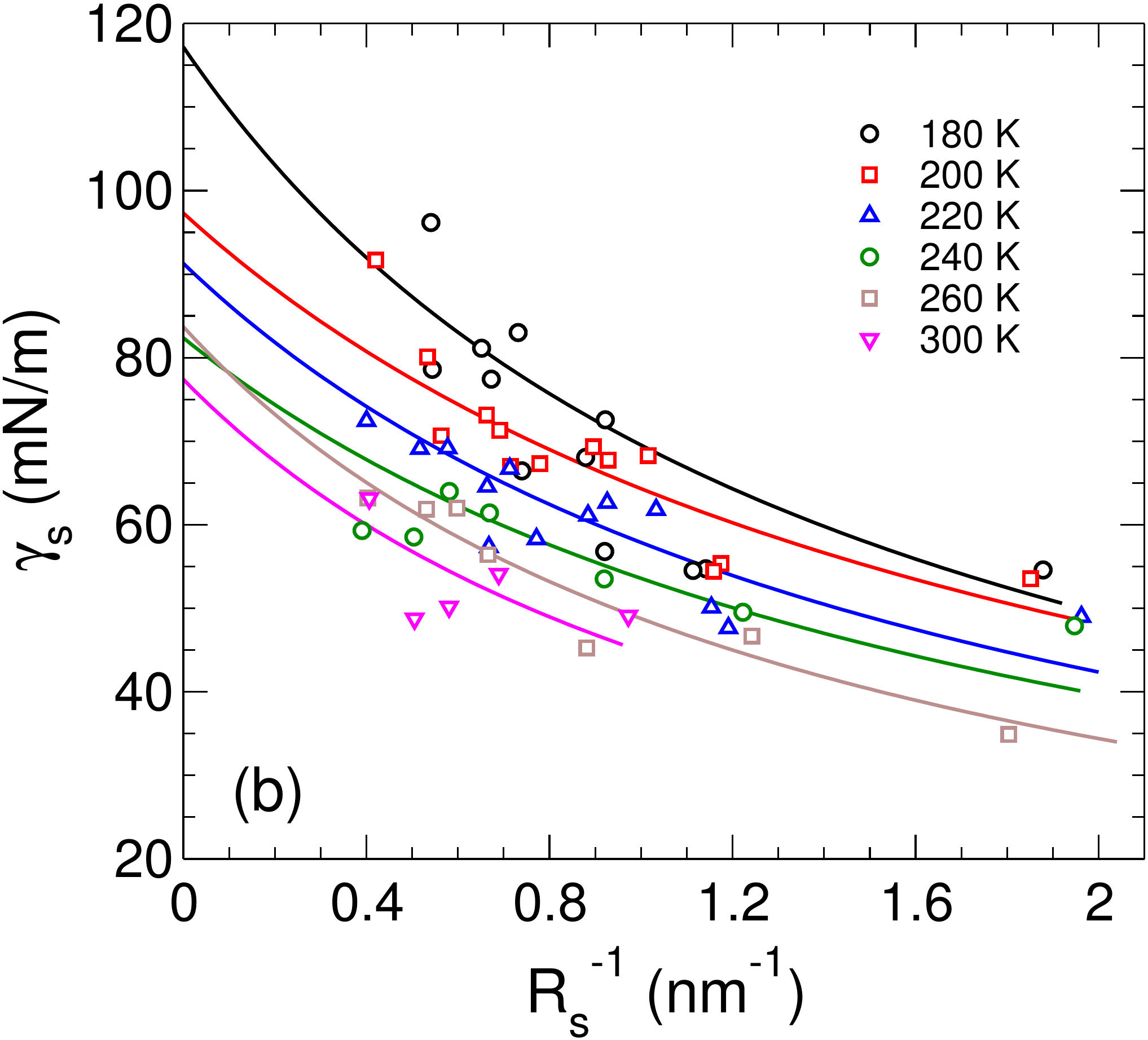}}
\caption[$\gamma_s$ as a function of $R_s^{-1}$.]{
$\gamma_s$ as a function of $R_s^{-1}$.  (a) $\gamma_s$ obtained from Eq.~\ref{gammastourqe} (symbols), where curves are fits to Eq.~\ref{tolman}. (b) $\gamma_s$ obtained from Eq.~\ref{gammastourqePn} (symbols), where curves are replotted from panel (a).
Curve intercepts estimate $\gamma_p$.
}
\label{gammaSfig}
\end{figure}

\begin{figure}[!htb]
\centerline{\includegraphics[scale=0.35]{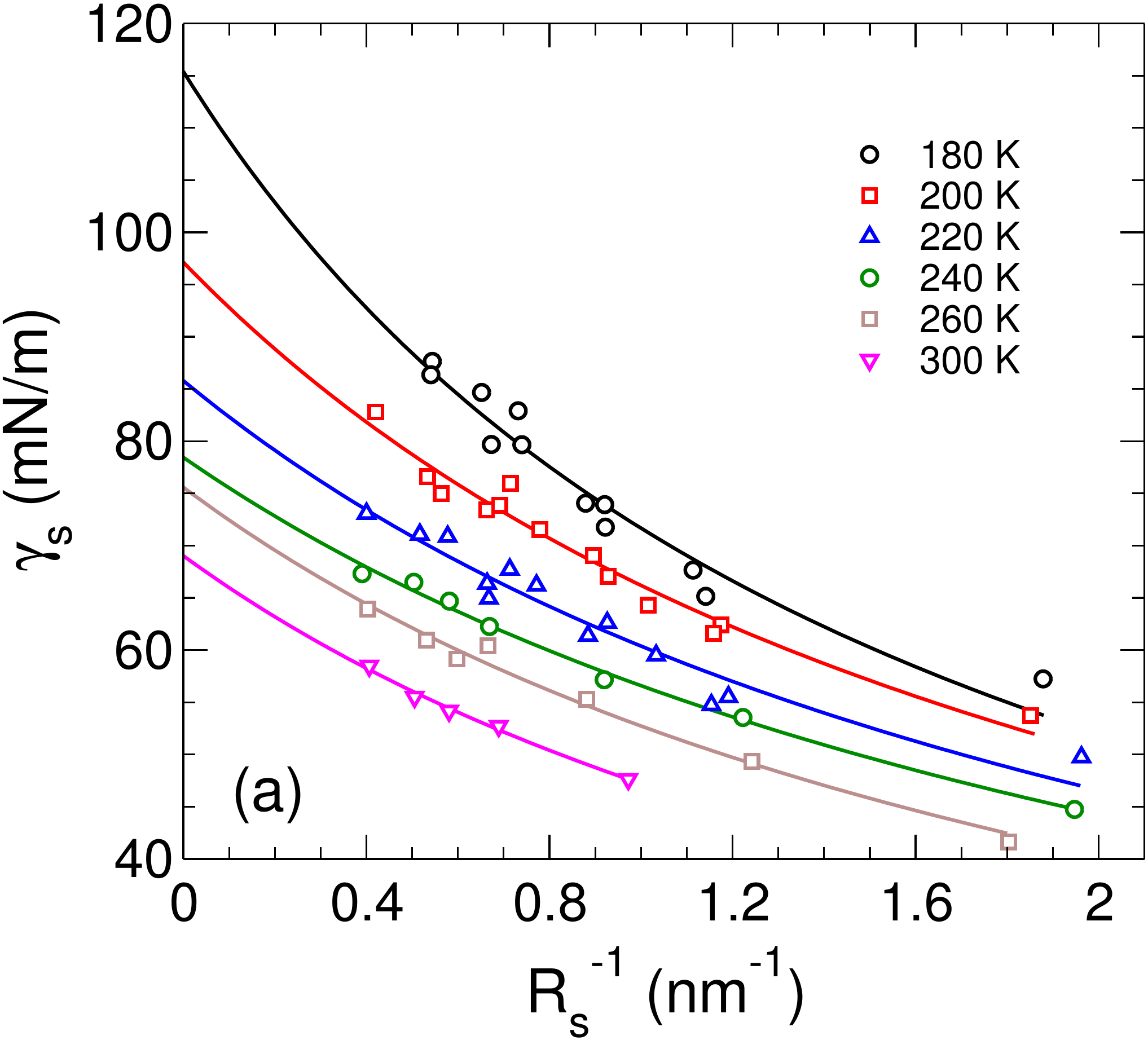}}
\centerline{\includegraphics[scale=0.35]{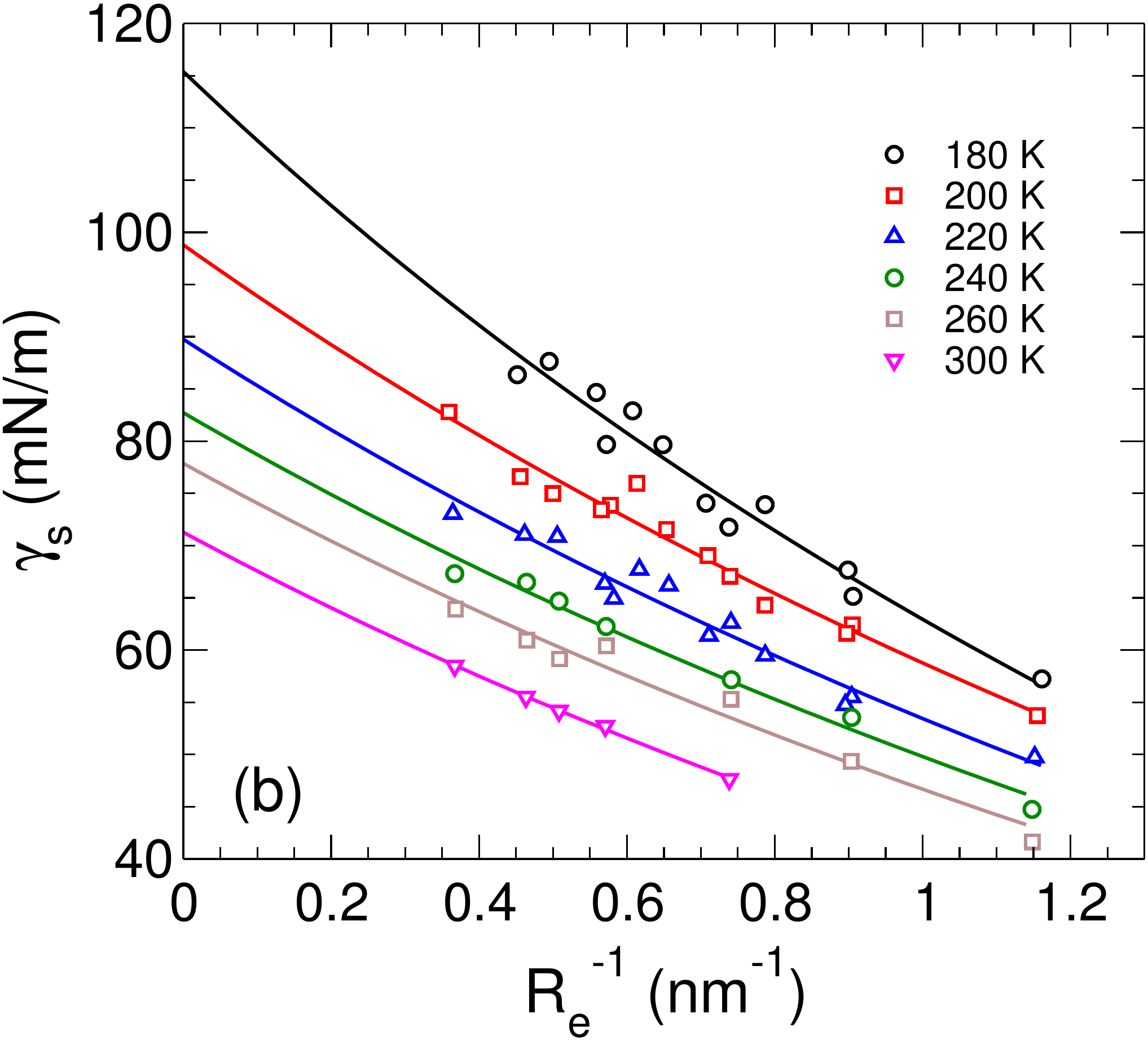}}
\caption[$\gamma_s$ as a function of $R_s^{-1}$ and $R_e^{-1}$.]{
$\gamma_s$ from Eq.~\ref{gammas3} as a function of (a) $R_s^{-1}$, with fits to Eq.~\ref{tolman} (solid lines). (b) $R_e^{-1}$, with fits to Eq.~\ref{tolmanRe} (solid lines).  Curve intercepts estimate $\gamma_p$.
}
\label{gammaS3fig}
\end{figure}

\begin{figure}
  \centering
\includegraphics[scale=0.25]{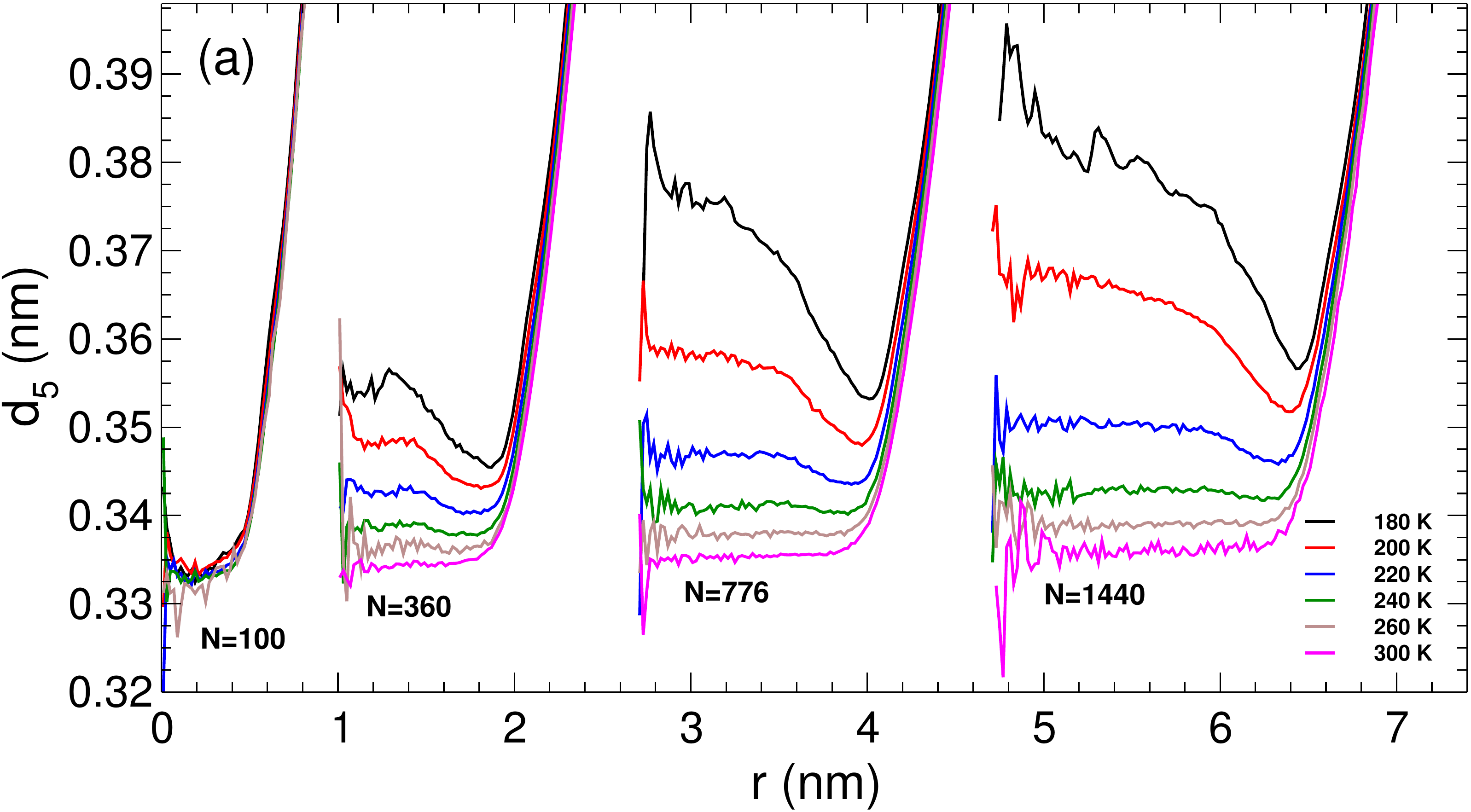}
\includegraphics[scale=0.25]{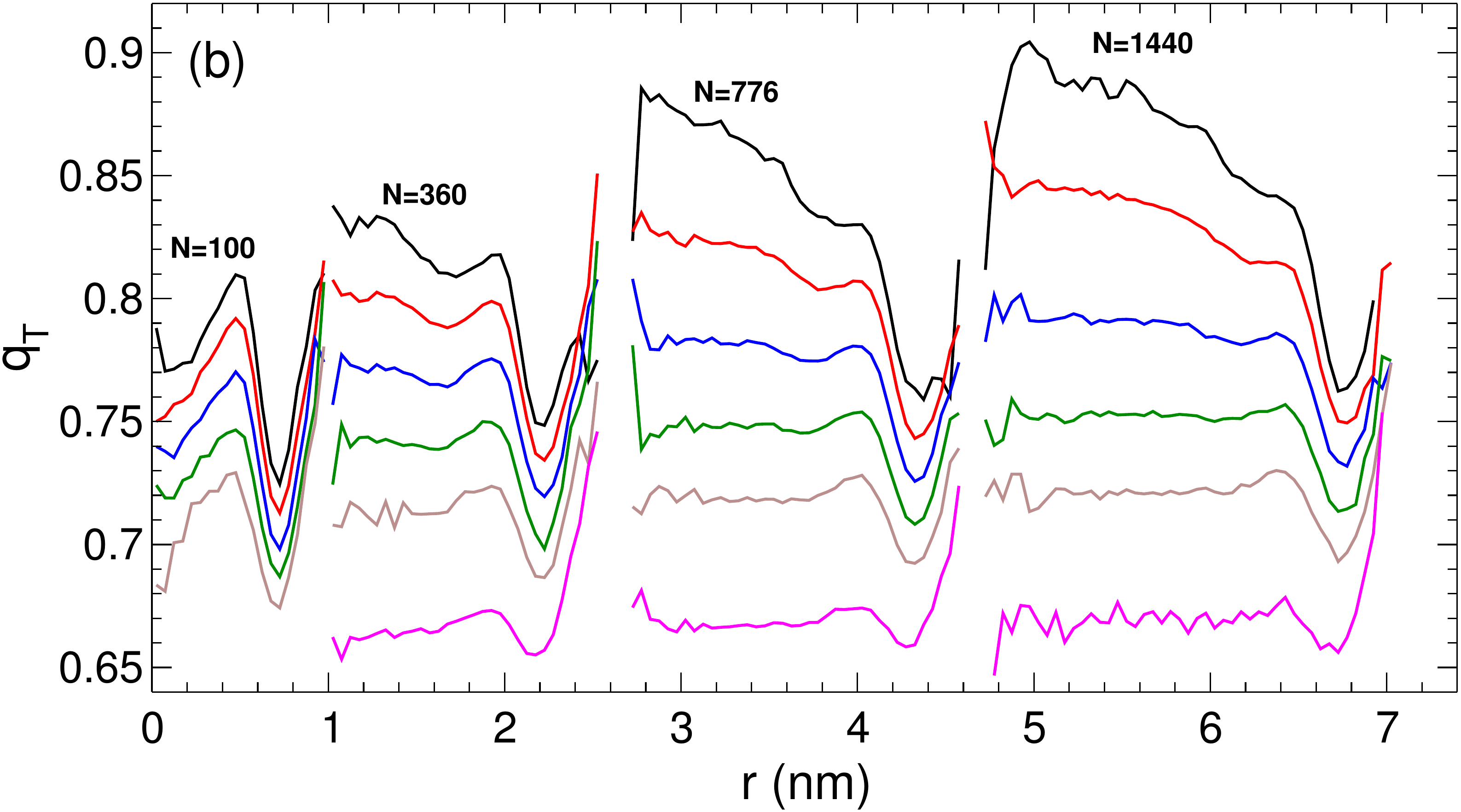}
\caption[$d_5$ and $q_T$ as a function of radius $r$.]{
(a) $d_5$ as a function of radius $r$ for various $N$ and $T$. The curves have been shifted horizontally by 1~nm for $N=360$, by 2.7~nm for $N=776$, and by 4.7~nm for $N=1440$. (b) $q_T$ as a function of radius $r$ for various $N$ and $T$. The curves have been shifted horizontally by 1~nm for $N=360$, by 2.7~nm for $N=776$, and by 4.7~nm for $N=1440$.    
}
\label{d5PROF}
\end{figure}

\begin{figure}[t]
  \centering
\includegraphics[scale=0.35]{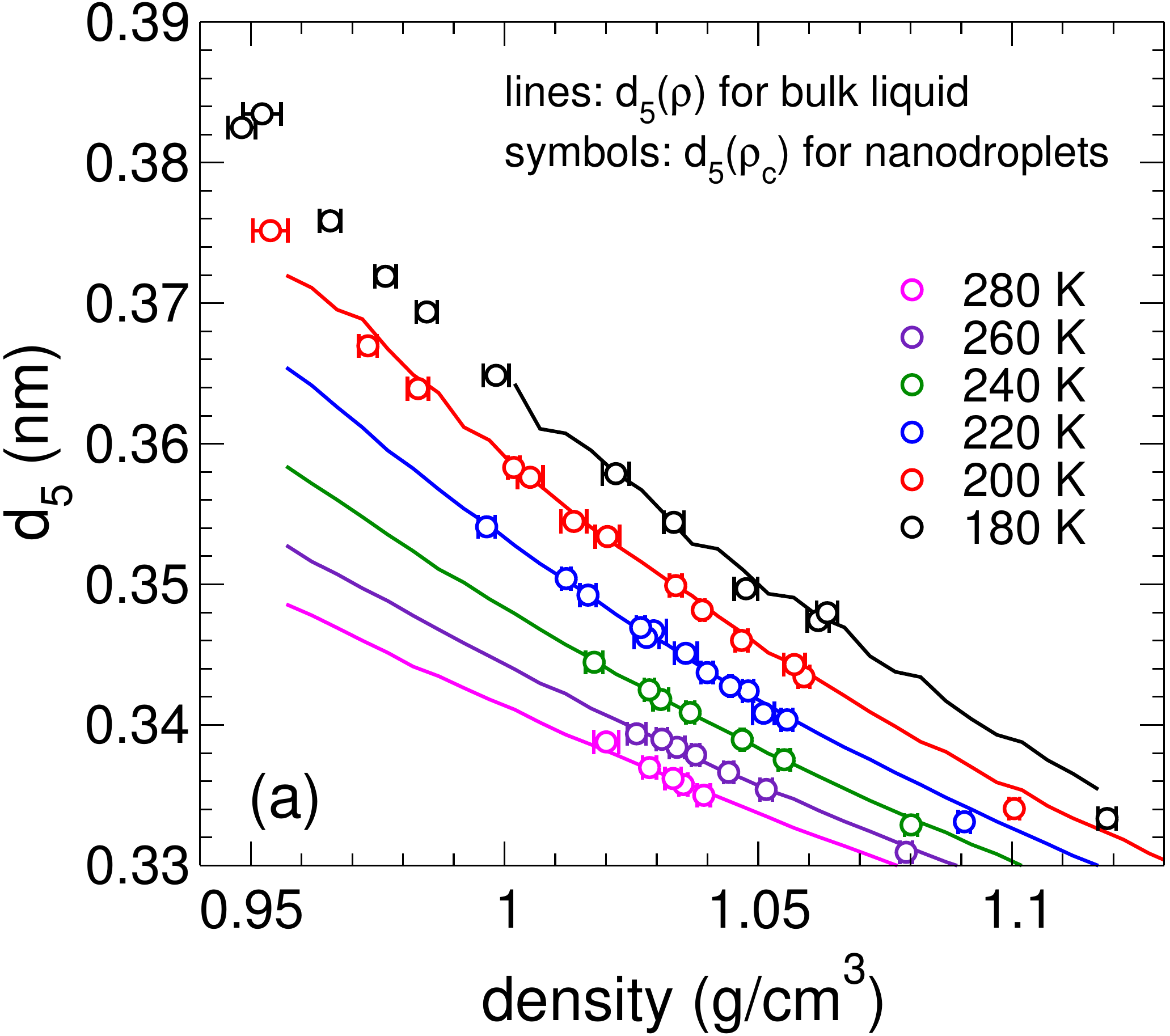}
\includegraphics[scale=0.35]{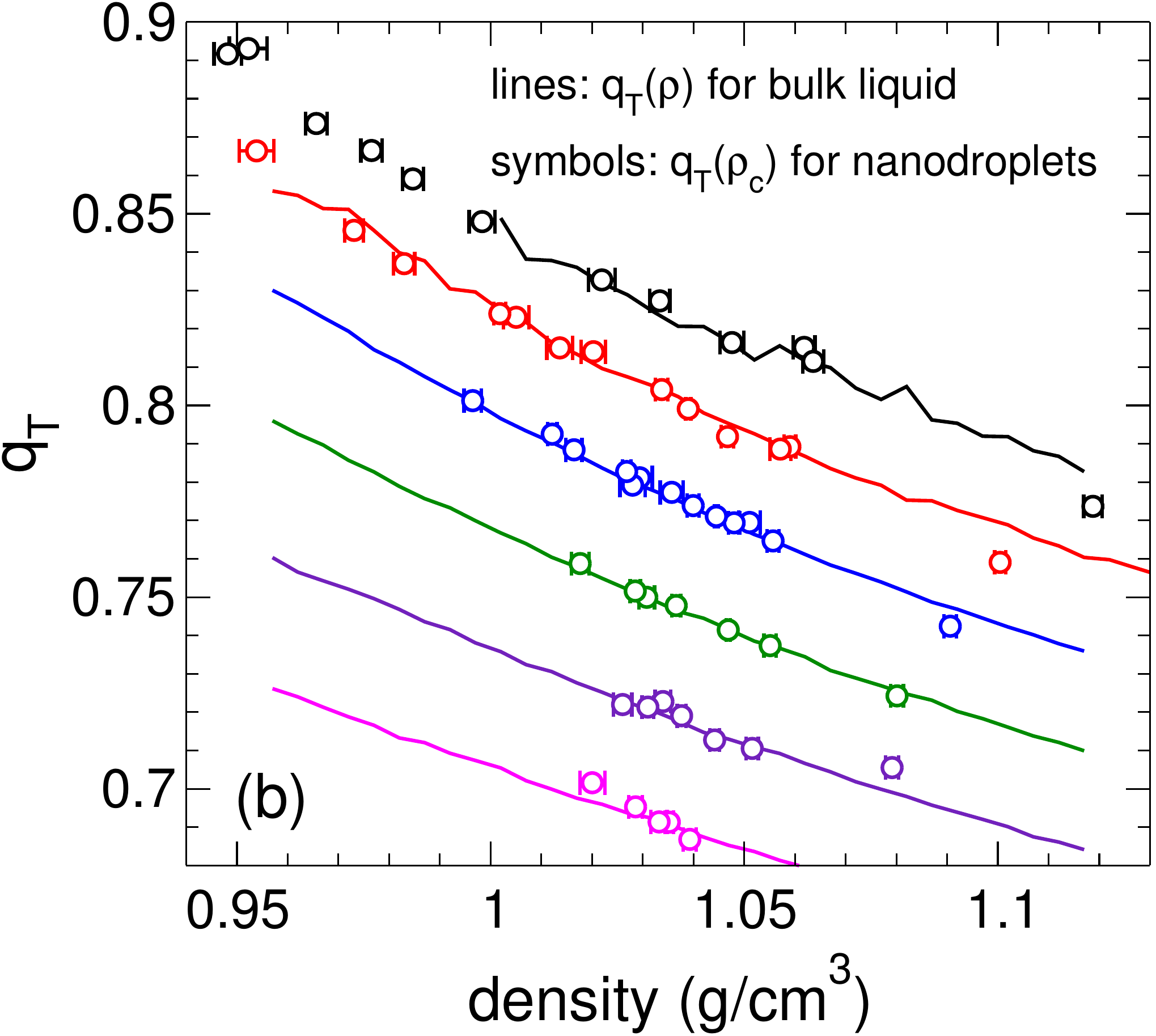}
\caption{Local measures of structure (a) $d_5$ and (b) $q_T$ as functions of density.
Symbols indicate data for nanodroplets while curves show results for bulk.  At $T=180$~K, it is difficult to
equilibrate the bulk liquid at low density.
}
\label{eosstruc}
\end{figure}

\section{Thermodynamic route}

In the thermodynamic route, we use the Young-Laplace equation in the form of  Eq.~\ref{young-laplaceRe} to determine $\gamma_p$ and $\delta$.
We equate $P_L$ with $\Delta P$, since the vapour pressure is negligible, and plot isotherms of $P_L$ as a function of $R_e^{-1}$ in Fig.~\ref{PLaplacefig}. The isotherms show that there is a significant pressure that naturally builds up in the interior of the droplets, and it can reach more than 200~MPa for $R_e^{-1}\simeq 1.2$~nm$^{-1}$ ($R_e\simeq 0.83$~nm). 

In Fig.~\ref{PLaplacefig} we study the curvature correction to $P_L$ as a function of $R_e$. Assuming $\delta=0$, the fits in Fig.~\ref{PLaplacefig}a  using Eq.~\ref{young-laplaceRe} (fitting only for $\gamma_p$) show that there is no obvious curvature correction to the Young-Laplace equation. To see how small $\delta$ is in our range of droplet sizes, we fit $P_L$ as a function of $R_e^{-1}$ at each $T$ with Eq.~\ref{young-laplaceRe} (fitting for both $\gamma_p$ and $\delta$), as shown in Fig.~\ref{PLaplacefig}b. We report the value of $\delta$ as a function of $T$ in Fig.~\ref{deltaPL},  and can discern no clear dependence of $\delta$ on $T$. The average small (positive) value of the Tolman length $\overbar{\delta}=0.055$~nm  explains the absence of strong curvature in the isotherms of Fig.~\ref{PLaplacefig}. 
As an alternative  way of obtaining $\gamma_p$ and $\delta$, we plot isotherms of $2/(P_LR_e)$ as functions of $R^{-1}_e$ in Fig.~\ref{PREfig}.
Since $\delta$ does not have an apparent dependence on $T$, we fit the isotherms in Fig.~\ref{PREfig} to Eq.~\ref{young-laplaceRev2} assuming a single common value of the fitting parameter $\delta$ for all $T$. As shown in Fig.~\ref{PREfig}, this global fit reasonably describes all the isotherms, and gives a value of $\delta = 0.056$~nm that is similar to $\overbar{\delta}$.  The intercepts in Fig.~\ref{PREfig} yield   $1/\gamma_p$ for each $T$.  As shown in Fig.~\ref{gammapfig}a, the values of $\gamma_p$ obtained in this way increase as $T$ decreases. 



In Fig.~\ref{gammapfig}a, we compare estimates for $\gamma_p$ assuming $\delta$=0 (obtained from the fits in Fig.~\ref{PLaplacefig}a) and $\delta \ne 0$ (obtained from the fits in Fig.~\ref{PLaplacefig}b).
At $T=180$~K the discrepancy  in $\gamma_p$ between assuming $\delta=0$ and $\delta\ne0$ appears to be outside of error, with the curvature-corrected result yielding a value of $\gamma_p$ approximately 10\% higher (blue squares versus 
red circles in Fig.~\ref{gammapfig}a).  For $T\ge 220$~K our estimates of $\gamma_p$ are also consistent with the extrapolation down to low $T$ of $\gamma_p$ obtained using the test-area method, taken from Eq.~6 in the work of Vega and de Miguel~\cite{Vega2007}.

To obtain $\gamma_s$ from the thermodynamics route, we combine Eqs.~\ref{young-laplace} and~\ref{delta} to obtain $\gamma_s=P_L \, (R_e-\delta)/2$, and use the values of $\delta$ for each $T$ obtained from fits shown in Fig.~\ref{PLaplacefig}b.  The resulting values of $\gamma_s$ are shown in Figs.~\ref{gammap-allTfig}a and \ref{gammap-300Kfig}.
%
%

\section{Mechanical route}

As discussed in Sec.~\ref{Intro}, $\gamma_p$ and $\delta$ can also be obtained using the mechanical route. To find $\gamma_p$ and $\delta$, we first evaluate $\gamma_s$ using Eq.~\ref{gammastourqe}, where we set $P^\alpha = P_L$   and $P^\beta=0$ since the vapour pressure in our simulations is negligible. In Fig.~\ref{gammaSfig}a we show isotherms of $\gamma_s$ as a function of $R_s^{-1}$, where $R_s$ is obtained from Eq.~\ref{Rsnumerical}. We see that $\gamma_s$ decreases with increasing $R_s^{-1}$ along isotherms, indicating that $\delta$ is positive. 
Fitting these isotherms with Eq.~\ref{tolman} yields curves 
from which $\gamma_p$ is estimated from the intercept at $R_e^{-1}=0$.  Our results for $\gamma_p$ obtained in this way are shown in Fig.~\ref{gammapfig}b.

Another way of evaluating $\gamma_s$ is through Eq.~\ref{gammastourqePn}. Isotherms of $\gamma_s$ from Eq.~\ref{gammastourqePn}  as a function of $R_s^{-1}$, where $R_s^{-1}$ is estimated using Eq.~\ref{RsnumericalPn}, are shown in Fig.~\ref{gammaSfig}b. Although the trend seems to indicate that $\gamma_s$ decreases with $R_s^{-1}$ using Eqs.~\ref{gammastourqePn} and~\ref{RsnumericalPn}, the noise resulting from subtracting $P_N(r)$ and $P_T(r)$ prevents useful fitting of $\gamma_s$. The solid curves shown in Fig.~\ref{gammaSfig}b are simply the fits taken from Fig.~\ref{gammaSfig}a, and show a general consistency between using Eqs.~\ref{gammastourqe} and \ref{Rsnumerical}, and using Eqs.~\ref{gammastourqePn} and \ref{RsnumericalPn}, with the former set suffering from less statistical scatter.

Unlike both Eqs.~\ref{gammastourqe} and~\ref{gammastourqePn}, which require the determination of $R_s$ to evaluate $\gamma_s$, Eq.~\ref{gammas3} does not involve calculating $R_s$. We plot $\gamma_s$ obtained from Eq.~\ref{gammas3} as a function of $R_s^{-1}$ in Fig.~\ref{gammaS3fig}a. We choose $R_s$ from Eq.~\ref{RsnumericalPn} because $\gamma_s$ in Eq.~\ref{gammas3} is derived from Eq.~\ref{gammastourqePn}. The absence of $R_s$ in Eq.~\ref{gammas3} seems to suppress the noise from $P_N(r)$. 
We fit the isotherms in Fig.~\ref{gammaS3fig}a to Eq.~\ref{tolman}, and we obtain values of
$\gamma_p$ and $\delta$ similar to those obtained from the isotherms in Fig.~\ref{gammaSfig}a,
as shown in Fig.~\ref{gammapfig}b and Fig.~\ref{deltaPL}.


To avoid any difficulty inherent in calculating $R_s$, another way of representing $\gamma_s$ is as a function of $R_e^{-1}$, as shown in Fig.~\ref{gammaS3fig}b. Regardless of which variant of the mechanical route is taken, we observe that $\gamma_s$ decreases as $R_e$ and $R_s$ decrease, $\delta$ is positive with little evidence for a dependence on $T$, and $\gamma_p$ increases as $T$ decreases.  The values of $\delta$ obtained from each variant of the mechanical route are shown for each $T$ in Fig.~\ref{deltaPL}.

%

\section{Comparison of thermodynamic and mechanical routes}

Fig.~\ref{gammapfig}b shows that the mechanical route yields values of $\gamma_p$ approximately 10\% larger than does the thermodynamic route, though the trends with $T$ are similar.
The mechanical route values for $\gamma_p$ are also generally higher than those from Vega and de Miguel~\cite{Vega2007}.
However, the best agreement with Ref.~\cite{Vega2007} between $T=240$~K and 300~K comes from using Eq.~\ref{gammas3} for calculating $\gamma_s$, Eq.~\ref{RsnumericalPn} for calculating $R_s$, and fitting the resulting isotherms (shown in Fig.~\ref{gammaS3fig}a) with Eq.~\ref{tolman} to obtain $\gamma_p$ and $\delta$.  

For all variants of both routes, $\gamma_p$ shows a striking departure from the extrapolated low $T$ behaviour presented in Ref.~\cite{Vega2007}.  The sharper than expected increase in $\gamma_p$ with decreasing $T$ occurs between 220 and 240~K, and is therefore consistent with crossing the Widom line at 230~K for bulk TIP4P/2005 at ambient pressure~\cite{vega2010}.  
In order for bulk properties of the liquid to influence $\gamma_p$ as obtained from nanodroplets, it is reasonable to expect that nanodroplet interiors are structurally similar to the bulk, an expectation for which we provide evidence in Section~VI.


For an independent comparison of $\gamma_s$, we show in Fig.~\ref{gammap-300Kfig} our results for  $\gamma_s$ as a function of $R_e$ as obtained from both the thermodynamic and mechanical routes at $T=300$~K along with the values from Lau {\it et al.}~\cite{Lau2015} obtained using the test-area method at $T=293$~K. We see that our results for $\gamma_s$ from the thermodynamic route are consistent with Lau, {\it et al}. However, the mechanical route gives significantly smaller values of $\gamma_s$. Smaller values of $\gamma_s$ and $R_s$ and larger values of $\delta$ for the mechanical route are also observed in nanodroplets interacting through the Lennard-Jones potential studied by Thompson, {\it et al}~\cite{Thompson1984}.

{\blue While Fig.~\ref{gammap-300Kfig} shows consistency in the value of $\gamma_s$ and its $R_e$ dependence between our thermodynamic route and the test-area method employed by Lau {\it et al.}~\cite{Lau2015}, other studies have found that $\gamma_s \approx \gamma_p$ for droplets as small as approximately 40 molecules  ($R_e \approx 0.6$~nm).  These studies employed techniques including excision of spherical portions from a bulk liquid~\cite{Samsonov2003}, a volume perturbation method allowing for a thermodynamic determination of the pressure tensor components~\cite{Ghoufi2011}, and a mitosis method by Joswiak et al.~\cite{Joswiak}.  Interestingly, Lau {\it et al.}~also carried out a mitosis method in a different study~\cite{Lau2015-2} (also finding that  $\gamma_s$ depends more weakly on $R_e$), and offered some discussion on the disparity between the mitosis and test-area methods.  All these other studies point to the validity of approximating $\gamma_s$ with $\gamma_p$ in estimating the Laplace pressure for very small nanodroplets.
}

To compare the difference in $\gamma_s$ as obtained from the mechanical and thermodynamic routes, we plot in Fig.~\ref{gammap-allTfig} isotherms of $\gamma_s$ as a function of $R_e$. Fig.~\ref{gammap-allTfig}b shows a significant change in $\gamma_s$ obtained from Eq.~\ref{gammas3} as droplet size varies. For a change in the nanodroplet radius from 1 to 3~nm, there is a 50\% increase in $\gamma_s$ at $T=180$~K, and 44\% at $T=300$~K. However, if we compare this with $\gamma_s$ estimated from the thermodynamic route
shown in Fig.~\ref{gammap-allTfig}a,
we  see that the isotherms are almost flat for $T\ge 220$~K, while there is only a 15\% difference in $\gamma_s$ across the droplet size range for $T\le200$~K. We also can see that $\gamma_s$ from the thermodynamic route is systematically larger than the mechanical route, which is once again consistent with Thompson,~{\it et~al}~\cite{Thompson1984}.

{\blue
As shown by Sampayo et al.~\cite{Sampayo2010}, disparity between thermodynamic and mechanical routes arises when energy fluctuations  (and not merely the average change in energy) become important in determining the free energy change when surface area is increased.  While for planar interfaces such fluctuations do not contribute significantly to the surface tension, the contribution for small droplets can be significant.  Thermodynamic routes include these fluctuations, while mechanical routes would require the addition of hypervirial terms to include the effects of fluctuations.  Ignoring these fluctuations in the mechanical route (i.e., when using pressure tensor components only) can lead, for example, to a change in sign in the determination of $\delta$ for Lennard-Jones droplets.  See also Ref.~\cite{Malijevsky2012} for discussion.
}

%

\section{Local structure ordering}


To quantify the structure of the interior of our water nanodroplets, we calculate the distance $d_5(r)$ between a molecule located at a distance $r$ from the centre of the droplet and its fifth-nearest-neighbour molecule (using distances between centres of mass).  A large value of $d_5(r)$ indicates that 
molecules tend to be four-coordinated, i.e. that the local tetrahedral network is well formed~\cite{ssp2001}.  

In Fig.~\ref{d5PROF}a we show $d_5(r)$ over a wide range of $N$ and $T$. We observe that $d_5(r)$ for droplet size $N=100$ is small and stays rather constant with $T$. 
The low value of $d_5$ indicates a collapse of the second neighbour shell around each molecule. This collapse is characteristic of the HDL form of water. 
The absence of any change in $d_5(r)$ with $r$ as we approach the surface indicates a disturbance in the tetrahedral network in the whole droplet. 
The overlap of the curves at different $T$ for $N=100$ suggests that droplets at this small size remain HDL-like both in the interior and at the surface regardless of how deeply we supercool them. 

As we increase the droplet size to $N=360$, the profiles systematically shift to higher value of $d_5$ in the interior as we cool to 180~K. This change is a signature of a crossover from HDL at high $T$ to LDL at low $T$. However, for $T\le 220$~K and $N=360$ in Fig.~\ref{d5PROF}a, there is a decrease in $d_5$ going from interior to surface, which indicates a disturbance of the tetrahedral network and an increase in density at the surface. For larger droplets, such as $N=776$, we see similar behaviour as for $N=360$, but the transformation spans a wider range of $d_5$. Moreover, for $N=776$ at $T=180$~K we see a monotonic decrease in $d_5$ as we approach the surface. This may reflect the emergence of structural transformation within the droplet.
The same scenario presents itself for $N=1440$.  At $T=180$~K, as $N$ increases from 100 to 1440, $d_5$ in the interior monotonically increases with $N$. This indicates that as $N$ increases, a better LDL forms in the interior of the droplets. 

To further probe the ordering inside the nanodroplets, we compute the local tetrahedral order parameter~\cite{qt},   {\blue 
\begin{eqnarray}
q_i &=& 1-\frac{3}{8}\sum_{j=1}^3\sum_{k=j+1}^4 \left[\cos\psi_{jik}+\frac{1}{3}\right]^2 \label{qiT}
\end{eqnarray}
}where $\psi_{jik}$ is the angle between 
{\blue an oxygen
atom $i$ and its nearest neighbour oxygen atoms $j$ and $k$. Subsequently, we define $q_T(r)$ as
the average value of $q_i$ for all molecules within a spherical shell bounded by radii $r\pm \Delta r/2$, where  $\Delta r=0.05$~nm.} 

%

We show how $q_T(r)$ changes  in Fig.~\ref{d5PROF}b. We see that $q_T$ is low for $N=100$ and it increases as we cool the droplet. Similar behaviour appears for $N=360$, 776, and 1440. However, for $T=200$ and 180~K, the increase in $q_T$ upon increasing $N$ becomes quite dramatic,  
supporting the suggestion that a better tetrahedral network forms as $N$ increases. 
For $N=1440$ at $T=180$~K, the core reaches 90\% of perfect tetrahedral order. 
The monotonic decrease in $q_T$ with $r$ for $N=776$ and $1440$ at $T=180$~K is consistent with the decrease with $r$ that we observe in $d_5$.

Our results for $d_5$ and $q_T$ suggest the progressive formation of LDL-like structure in the interior of our droplets as $T$ decreases.  The structural results presented here are also consistent with the evolution of the density profiles of our droplets presented in Ref.~\cite{NATURECOMM}.  The transformation from HDL to LDL in the droplet interior can also explain the change in behaviour in $\gamma_p$ for $T < 240$~K shown in Fig.~\ref{gammapfig}.  Since LDL is a more structured liquid than HDL, with a better formed hydrogen bond network, we expect that the interface between LDL and the vapour phase will have a higher surface tension than for the interface between HDL and the vapour.

To illustrate the structurally bulk-like character of our droplet interiors, we plot $d_5$ and $q_T$ as functions of density in Fig.~\ref{eosstruc}. To compute the density, we define the density within the core of our droplets as $\rho_c=m\langle {\cal N}/ {\cal V} \rangle$, where ${\cal N}$ is the number of O atoms within a defined core radius $r_c=0.5$~nm of the droplet centre, ${\cal V}$ is the total volume of the Voronoi cells for these atoms~\cite{voronoi}, and $m$ is the mass of a water molecule.
Since in the smallest droplets surface effects extend closer to the centre of droplet, we use $r_c=0.25$~nm for $N\le 205$. 
{\blue Similarly, we define $d_5$ and $q_T$ for droplet interiors by averaging the corresponding local quantities for particles within $r_c$ of the droplet center (molecules for $d_5$ and O atoms for $q_T$).} 
Fig.~\ref{eosstruc} shows the agreement between $d_5$ and $q_T$ as functions of density for bulk systems and droplets. 
This correspondence demonstrates that the core of the droplets for our range of $N$ is bulk-like. 
These structurally bulk-like interiors are consistent with
the possibility that our extrapolated values of $\gamma_p$, obtained from the behavior of nanoscale droplets, approximate those for bulk planar liquid-vapour interfaces, and hence that the anomalous increase in $\gamma_p$ we observe below 230~K reflects the bulk liquid anomalies associated with crossing the Widom line of the LLPT.

\section{Discussion and conclusions}

We estimate the surface tension of water nanodroplets using the TIP4P/2005 model over a wide range of $N$ and $T$. We do so from an evaluation of the components of the pressure tensor inside the droplets~\cite{NATURECOMM} using the method described in Ref.~\cite{malek2}. From the pressure tensor components, we determine the isotropic pressure $P_L$ in the interior of the droplets. 
This allows us to calculate the surface tension with two approaches: using the Young-Laplace equation directly, and using the variation of the pressure tensor components with distance from the droplet center.
The direct route, which we call the thermodynamic route, requires $P_L$ and $R_e$ to estimate $\gamma_s$, $\gamma_p$ and $\delta$ as fit parameters, and the mechanical route  evaluates $\gamma_s$ and $R_s$ from the pressure tensor components, and yields  $\gamma_p$ and $\delta$ from fitting.
{\blue It should be noted, however, that the analysis carried out by Gibbs (see e.g.~Ref.~\cite{Rowlinson1982}) and reiterated by Tolman~\cite{Tolman1949}, imply that the interior pressure that should be used in Eq.~\ref{young-laplace} is that of the bulk fluid with the same chemical potential as the droplet interior.  It would be interesting to quantify the differences in the calculated surface tension that arise from using this definition instead of the directly-calculated pressure, particularly for smaller droplets where differences may be significant.}


Isotherms of $P_L$ plotted as a function of $R_e^{-1}$ on the assumption that the surface of tension acts at $R_e$ (i.e. that $\delta=0$) show a linear dependence between $P_L$ and $R_e^{-1}$ that is valid for droplets as small as 0.86~nm in radius. To validate this apparent linearity, we insert the Tolman length correction into the Young-Laplace equation and find that $\delta$ is positive and small with a value of  $0.055\pm 0.021$~nm. Moreover, $\gamma_p$ values for $T\ge 220$~K from this thermodynamic route, regardless of whether we assume $\delta$ is zero or not, 
are consistent with the extrapolation of $\gamma_p$ obtained for TIP4P/2005 using the test-area method~\cite{Vega2007}, a thermodynamic method, 
as shown in Fig.~\ref{gammapfig}a. 

We compute $\gamma_p$ from the mechanical approach by first finding $\gamma_s$ and $R_s$ using Eqs.~\ref{gammastourqe} and~\ref{Rsnumerical}; then by 
using Eqs.~\ref{gammastourqePn} and~\ref{RsnumericalPn}, which produces consistent, but noisier results; 
and finally by using Eqs.~\ref{gammas3} and~\ref{RsnumericalPn}. For our range of $T$ and $N$, we show that $\gamma_s$ decreases as $R_s$ decreases. 
Fitting these results with Eq.~\ref{tolman} results in positive and rather large values of $\delta=0.32\pm 0.02$~nm from Fig.~\ref{gammaSfig}a, and $\delta=0.21\pm 0.01$~nm from Fig.~\ref{gammaS3fig}a. Although these two values do not overlap within error, they both suggest that $\delta$ from the mechanical route is significantly larger than the value from the thermodynamic route.  
Moreover, estimates of $\gamma_p$ obtained from fitting mechanical-route results tend to be higher than thermodynamic-route results, as apparent in Fig.~\ref{gammapfig}b. 
However, if we consider $\gamma_s$ from Eq.~\ref{gammas3} as a function of $R_s^{-1}$ as shown in Fig.~\ref{gammaS3fig}a, the $\gamma_p$ values resulting from fitting  with Eq.~\ref{tolman} are consistent with the thermodynamic route and with Vega and de Miguel's extrapolation for $T\ge 240$~K.

We also conclude that $\gamma_s$ from the thermodynamic route remains relatively constant as we vary $R_e$ for $T\ge 220$~K, but shows larger variation at $T=200$ and 180~K, where it changes by  15\% over the range of droplet sizes we use. In contrast, $\gamma_s$ from the mechanical route increases significantly with $R_e$, resulting in almost a 50\% change in $\gamma_s$ at $T=180$~K.  These results are equivalent to 
$\delta$ being small for the thermodynamic route and large for the mechanical route.  

At 300~K, our thermodynamic results for $\gamma_s$ as a function of droplet size are consistent with those of Lau,~{\it et al.}~\cite{Lau2015}, while those from the mechanical route are not.  One might conclude, therefore, that the mechanical route for determining $\gamma_s$ and $\delta$ lacks validity, and the relatively large value of $\delta=0.2$ -- 0.3~nm should be rejected in favour of the smaller value of $\delta\approx0.06$ determined from the thermodynamic route.  However, as $\delta$ is the difference between $R_e$ and $R_s$, which is understood to be where the surface tension acts, values in the range of 0.2 to 0.3 nm are reasonable given the locations of $R_e$ and the negative pressure minima in Fig.~\ref{pressprofiles}.  In sum, our work confirms the discrepancy between the mechanical and thermodynamic routes that has been previously noted in the literature, and so 
supports the need for a better theoretical understanding of the connection between the two.


The marked increase in $\gamma_p$ 
for $T<220$, as shown in Fig.~\ref{gammapfig}, approximately coincides with the crossing of the Widom line at $T=230$~K for bulk TIP4P/2005 water at ambient pressure~\cite{vega2010}, 
and hence, is correlated to the LLPT occurring in this water model.  This increase in $\gamma_p$ is consistent across both the mechanical and thermodynamic routes.  
Our results thus confirm the scenario predicted theoretically in Refs.~\cite{feeney,hruby2004,hruby2005}, in which $\gamma_p$ increases more rapidly with decreasing $T$ when the system enters the $T$ regime below the Widom line where LDL-like properties begin to dominate the bulk behavior.  We also note that Ref.~\cite{hruby2004} predicts that the surface of a deeply supercooled water nanodroplet will exhibit a dense surface layer relative to the bulk-like density of the droplet interior.  This prediction is confirmed by the density profiles presented in Ref.~\cite{NATURECOMM}, and is consistent with the radial variation of the structural properties presented in Section~VI.
{\blue The sudden increase in $\gamma_p$ at low $T$ that we infer from our droplet simulations was also observed in simulations of planar interfaces using the WAIL potential for water and was also interpreted as evidence for the LLPT scenario~\cite{Rogers2016}.}

Characterizing how local structure varies with radial distance from the center of the droplet with $d_5$ and $q_T$, we see behavior consistent with the formation of a well-ordered random tetrahedral network at low $T$ and large $N$ within droplet interiors.  Furthermore, the dependence of these structural measures on local density match that of bulk TIP4P/2005 water.  Hence, from a structural perspective, the interiors of our nanodroplets are characteristic of the bulk.


We conclude that $\gamma_s$ and $R_s$ determined from the mechanical route are smaller than the values evaluated in the thermodynamic route, leading to larger values of $\delta$ and  $\gamma_p$. However, both routes give a positive value of $\delta$ for our range of $T$ and $N$, and suggest that $\delta$ is independendent of $T$. Moreover, assuming the validity of thermodynamic route, for $R_e\ge 1$~nm we can ignore the curvature correction and use the planar surface tension to estimate the Laplace pressure inside water nanodroplets to within 15\% down to 180~K.    This last point is of practical importance for the estimation of the interior pressure in real water nanodroplets, for which the Laplace pressure is not easily measured directly.

\begin{acknowledgements}
ISV and PHP thank NSERC for support.  
PHP also acknowledges support from the Dr.~W.~F.~James Research Chair Program.
Computational resources were provided by ACENET and Compute Canada.  
\end{acknowledgements}

\end{document}